\documentclass{aa}  
\usepackage{natbib}
\usepackage{graphicx}
\usepackage{amssymb}
\usepackage{epsfig}
\begin{document}
\def\sizex{16.0 cm}
\def\bigx{10.0 cm}
\def\smallerxsize{7.0 cm}
\def\smallxsize{10.0 cm}
\def\smallysize{12.0 cm}

\title {The VIRMOS deep imaging survey II: \\ CFH12K BVRI optical data
  for the 0226-04 deep field\thanks{Based on observations obtained
  at the Canada--France--Hawaii Telescope (CFHT) which is operated by
  the National Research Council of Canada, the Institut des Sciences de
  l'Univers of the Centre National de la Recherche Scientifique and the
  University of Hawaii.}}
\titlerunning{CFH12K BVRI optical data for the 0226-04 deep field}

\author{H.\ J.\ McCracken \inst{1,2,7} \and M.  Radovich \inst{3,5} \and
  E. Bertin\inst{3,4} \and Y. Mellier \inst{3,4} \and M. Dantel-Fort \inst{4} \\
  O.  Le F\`evre \inst{2} \and J. C.  Cuillandre \inst{6} \and S.  Gwyn
  \inst{2} \and S. Foucaud \inst{2} \and G. Zamorani \inst{7}}

\offprints {H.\ J.\ McCracken}
%\date{\fbox{\sc Draft Version:\today}}
\date{}
\institute{University of
  Bologna, Department of Astronomy, via Ranzani 1, 40127 Bologna, Italy
  \and Laboratoire d'Astrophysique de Marseille, Traverse du Siphon,
  13376 Marseille Cedex 12, France \and Institut d'Astrophysique de
  Paris, 98 bis Boulevard Arago, 75014 Paris, France \and Observatoire
  de Paris, LERMA, 61 Avenue de l'Observatoire, 75014 Paris, France
  \and Osservatorio di Capodimonte, via Moiariello 16, 80131 Napoli,
  Italy \and Canada-France-Hawaii Telescope, 65-1238 Mamalahoa Highway, Kamuela, HI 96743 \and Osservatorio Astronomico di Bologna, Via Ranzani 1, 40127 Bologna, Italy}
 
\abstract{In this paper we describe in detail the reduction,
  preparation and reliability of the photometric catalogues which
  comprise the CFH12K-VIRMOS deep field. This region, consisting of
  four contiguous pointings of the CFH12K wide-field mosaic camera,
  covers a total area of $1.2\deg^2$. The survey reaches a limiting
  magnitude of $B_{AB}\sim 26.5$, $V_{AB}\sim26.2$, $R_{AB}\sim25.9$
  and $I_{AB}\sim25.0$ (corresponding to the point at which our
  recovery rate for simulated point-like sources sources falls below
  $50\%$).  In total the survey contains 90,729 extended sources in the
  magnitude range $18.0<I_{AB}<24.0$.  We demonstrate our catalogues
  are free from systematic biases and are complete and reliable down
  these limits. By comparing our galaxy number counts to previous
  wide-field CCD surveys, we estimate that the upper limit on
  bin-to-bin systematic photometric errors for the $I-$ limited sample
  is $\sim 10\%$ in this magnitude range. We estimate that $68\%$ of
  the catalogues sources have absolute per co-ordinate astrometric
  uncertainties less than $\vert\Delta\alpha\vert\sim0.38\arcsec$ and
  $\vert\Delta\delta\vert\sim0.32\arcsec$. Our internal
  (filter-to-filter) per co-ordinate astrometric uncertainties are
  $\vert\Delta\alpha\vert\sim0.08\arcsec$ and
  $\vert\Delta\delta\vert\sim0.08\arcsec$. We quantify the completeness
  of our survey in the joint space defined by object total magnitude
  and peak surface brightness.  We also demonstrate that no significant
  positional incompleteness effects are present in our catalogues to
  $I_{AB}<24.0$. Finally, we present numerous comparisons between our
  catalogues and published literature data: galaxy and star counts,
  galaxy and stellar colours, and the clustering of both point-like and
  extended populations. In all cases our measurements are in excellent
  agreement with literature data to $I_{AB}<24.0$. This combination of
  depth and areal coverage makes this multi-colour catalogue a solid
  foundation to select galaxies for follow-up spectroscopy with VIMOS
  on the ESO-VLT and a unique database to study the formation and
  evolution of the faint galaxy population to $z\sim1$ and beyond.
   }

\maketitle
\section{Introduction}
\label{sec:introduction}

Deep multi-colour observations remain one of the simplest and most
useful tools at our disposal to investigate the evolution and
properties of the faint galaxy population. The first digital studies of
the distant Universe \citep{1991MNRAS.249..498M,1991ApJ...369...79L,TS}
pushed catalogue limiting magnitudes beyond the traditional $B_J=24$
boundary of previous photographic works.  They provided the first
glimpse of the Universe at intermediate redshifts and tentative upper
limits on the numbers of $z>3$ galaxies \citep{1990ApJ...357L...9G}.
Following these studies, thousand-object spectroscopic surveys in the
mid-nineties \citep{1996MNRAS.280..235E,1995ApJ...455...50L}, combined
with reliable photometry provided by charge-coupled devices (CCDs)
mounted on 4m telescopes, sketched out for the first time the broad
outlines of galaxy and structure evolution for normal $L^*$ galaxies in
the redshift range $0<z<1$
\citep{1996ApJ...461..534L,1995ApJ...455..108L}.

At higher redshifts, the Hubble Deep Fields project \citep{WBD} has
demonstrated just how much can be accomplished beyond the spectroscopic
limit of even 10m telescopes, provided photometric observations are
sufficiently accurate. And in the last five years, searching for the
passage of the Lyman break through broad-band filters has proved to be
an extremely efficient way to locate objects at $z>3$ and above
\citep{1996ApJ...462L..17S,MFD,1993AJ....105.2017S}. These recent
studies have provided a unique insight into the formation of
large-scale structures and the relationship between mass and light in
the regime where theoretical predictions depend sensitively on the
underlying model assumptions \citep{1998ApJ...505...18A}.  Meanwhile,
at lower redshifts, the 2dF, 2MASS and Sloan surveys
\citep{2001MNRAS.328.1039C,2000AJ....120..298J,2000AJ....120.1579Y} are
now providing us with a detailed picture of local Universe and a solid
reference point for studies of the distant Universe.

To progress beyond these works at $z\sim1$ demands imaging catalogues
at least an order of magnitude larger than previously available from
surveys based on single CCD detectors. One simple reason is that many
objects of interest are \textit{rare}; surface densities of Lyman-break
galaxies, for instance, are around $\sim1~$arcmin$^{-2}$ at
$I_{AB}<24.0$; amassing large samples of such galaxies requires
coverage of a correspondingly large area of sky. In the recent years,
large format multiple-CCD cameras such as the University of Hawaii's 8K
camera (UH8K; \cite{1994SPIE.2198..810L}), containing many separate CCD
detectors and with a field of view of around $\sim 0.25$~deg$^2$ have
become available and several surveys are now underway or have been
recently completed using these cameras.  For example, the Canada-France
Deep Fields Survey (CFDF) covered three survey fields of the
Canada-France Redshift Survey \citep[CFRS;][]{1995ApJ...455...50L} with
the UH8K camera, providing $UBVI$ photometry for around $50,000$
objects \citep{2001A&A...376..756M}.  With even larger, more efficient
instruments such as the Canada-France-Hawaii Telescope's 12K mosaic
camera (CFH12K) it has now become feasible to conduct surveys
subtending degree scales on the sky and reaching mean redshifts of
$z\sim1$.  Furthermore, the advent of very high throughput
spectrographs such as VIMOS \citep{2000SPIE.4008..546L} demands a new
generation of wide-area deep digital survey catalogues in which
systematic errors are rigorously controlled. It is also important to
develop robust and reliable tools which can effectively deal with the
even larger volumes of data which will be produced by the forthcoming
500-night Canada-France-Hawaii-Telescope Legacy Survey (CFHTLS) project
which will use the new $1\deg^2$ field-of-view MEGACAM instrument
\citep{2000SPIE.4008..657B}.

In this paper we describe $BVRI$ photometric observations which have
been carried out in the VIRMOS-VLT deep field, a $1.2~\deg^2$ high
galactic latitude field of low galactic extinction which will be the
centrepiece of the VIRMOS-VLT deep spectroscopic survey
\cite[VVDS;][]{2001defi.conf..236L}. Several tens of thousands of
galaxy spectra will be acquired here to a limiting magnitude of
$I_{AB}<24.0$. Observations described in this paper were made using the
CFH12K mosaic camera: the broad spectral response and sensitivity of
12K, compared to previous-generation cameras such as the UH8K, make it
an ideal instrument for this kind of survey (the DEEP2 redshift survey
\citep{1998wfsc.conf..333D} are also using the CFH12K camera for survey
pre-selection).  In addition to the core $BVRI$ dataset described here,
near-ultraviolet, near-infrared, X-ray and 1.4 GHz radio observations
have also been carried out in this field, and will be presented in
forthcoming papers.  A companion paper (Le F\'evre et al., paper I)
describes the overall strategy of the survey.  Finally, a third paper
will describe shallower optical data. This comprises three $4~\deg^2$
fields (the fields at 14hrs, 22hrs and 10hrs) and a $2~\deg^2$
extension to the field described here. In each of these fields spectra
will be acquired from a sample of galaxies magnitude limited at
$I_{AB}=22.5$.

This paper is organised as follows. In Section~\ref{sec:obs} we
describe observations which were carried out;
Section~\ref{sec:data-reduct} describes how this data was reduced and
includes details of how both astrometric and photometric solutions were
computed to produce the final output images. In
Section~\ref{sec:catal-prep} we explain how object catalogues were
produced from these images.  In Section~\ref{sec:data-qual-assessm} we
present a detailed quality assessment of the catalogues, where
comparisons are made to existing deep catalogues. Our conclusions and
summary are presented in Section~\ref{sec:summary}.

\section{Observations} 
\label{sec:obs}

This paper describes the F02 deep field of the VIRMOS-VLT survey. This
field is centred at $02^h 26^m 00^s$ $-04^{\circ} 30' 00''$ (J2000)
which corresponds to a galactic latitude and longitude of
$(172.0,-58.0)$ respectively. All observations described here were
carried out using the CFH12K camera over a series of three observing
runs from November 1999 to October 2000.  The CFH12K is a 12,288 by
8192 pixel mosaic camera comprising twelve $2048\times 4096$ 15 micron
pixel thinned backside illuminated MIT Lincoln Laboratories CCID20
devices.  It has a total field of view of $42\arcmin\times28\arcmin$
\citep{2000SPIE.4008.1010C}. The average gain over all CCDs is
$1.6e^-$/ADU. The pixel scale of the detector is approximately
$0.205\arcsec/$pixel. The CFHT wide-field corrector produces a radial
distortion amounting to $\sim15$ pixels at the camera field of view
\citep{1996PASP..108.1120C} which necessitates resampling our images
before stacking. In total, the F02 deep field comprises four pointings
of CFH12K in four filters covering a total area of $1.2\deg^2$.

\begin{figure}
\resizebox{\hsize}{!}{\includegraphics{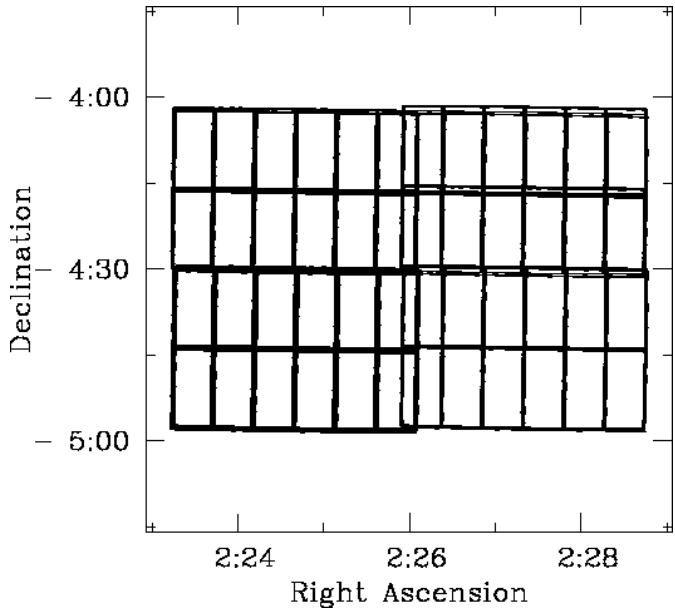}}
  \caption{An outline of all CCD frames contained in the $I-$ band stack. 
    Shown in this Figure are four separate pointings of the CFH12K
    camera. At each pointing there are 7-8 individual exposures
    comprising 12 CCDs in each. At the scale of this plot, it is not
    possible to see the displacement caused by the dither pattern
    offsets.}
  \label{fig:ditherplot}
\end{figure}

Cosmetic quality of the detectors is generally good, with less than 200
bad columns in total, most of which are concentrated on a single
detector. In addition, CFHT provides pixel masks which allow these bad
pixels to be easily removed.  The average gap between the charge
coupled devices (CCDs) in the east west-direction is $7.8''$, whereas
the north-south gap is $6.8''$. To fill these gaps in the final stacked
mosaic, exposures were arranged in a dither pattern of radius $15''$.
Figure~\ref{fig:ditherplot} shows the layout of all $I-$band
observations made on the field. Each rectangular box corresponds to the
outline of one CCD image. At the scale of this plot it is not possible
to see the dither pattern displacements.

\begin{figure}
\resizebox{\hsize}{!}{\includegraphics{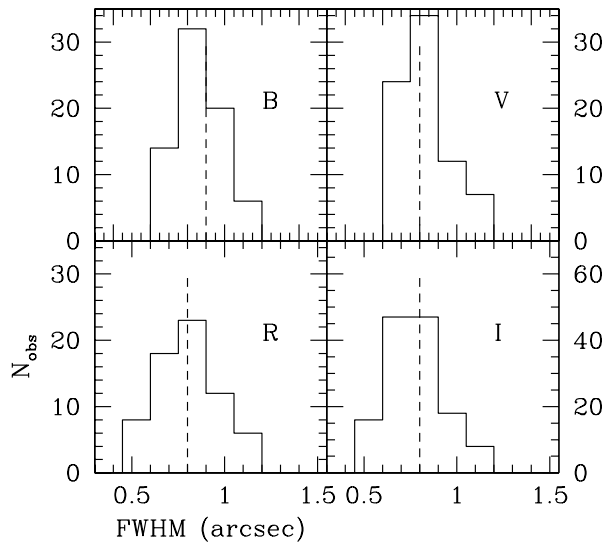}}
  \caption{Seeing distribution for CCD images used in the F02 mosaic
    for each of the four filters. The dashed line shows the median
    seeing; note that in all cases this is less than $1''$. All images
    with a seeing higher than $1.2''$ were discarded and are not shown
    in this Figure.}

  \label{fig:seeingplot}
\end{figure}

Exposure times per image in each filter were typically 1800s for $B$,
$1080s$ for $V$, $1200s$ for $R$ and $720s$ for $I$.  The total
exposure times in each of the four filters at each of the four
pointings are listed in Table~\ref{tab:vvds.fields}.  In
Figure~\ref{fig:seeingplot} we show the distribution of full-width
half-maximum (FWHM) seeing values for all the input images. All data
with FWHM values greater than $1.2''$ were discarded.

\begin{table*}
\begin{tabular}{*{8}{c}}

{\bfseries Field}     & R.A. (2000) & Dec. (2000) & Band & Exposure time &
Median seeing  & Completeness limit\\
                      &             &             &      &    (P1,P2,P3,P4)(hours)    &
(arcsec) & ($AB$ mags)           \\
\hline                                                                                                                  
                      &             &             &      &               &
&                       &           \\
{\bfseries 0226-0430} & 02:26:00    & $-$04:30:00 & B    &      6.0, 5.5, 5.5, 5.0      &
0.9    &        26.5          \\
                      &             &             & V    &      4.0, 7.0, 7.0, 4.0      &
0.8    &        26.2          \\
                      &             &             & R    &      3.8, 3.1, 3.7, 3.3      &
0.8    &        25.9          \\
                      &             &             & I    &      3.8, 3.2, 3.0, 2.8      &
0.8    &        25.0          \\
\end{tabular}
\caption{Details of the images used. We list the total
  integration time in each of the four pointings and the completeness
  limit of each filter, based on simulations described fully in
  Section~\ref{sec:limit-magn-catal}.}
\label{tab:vvds.fields}
\end{table*}

\section{Data reductions}
\label{sec:data-reduct}

Reducing observations taken by mosaic camera detectors present unique
challenges in terms of the quantity of data which must be processed.
For the project described in this paper, several thousand separate CCD
images were processed totalling several hundred gigabytes of data; at
least double this quantity of disk space is needed to combine these
images into a single mosaic. The very large volume of data involved
meant that all processing was carried out at TERAPIX, a data reduction
centre at the Institut d'Astrophysique de Paris designed specially to
process large quantities of wide-field mosaic camera data. One of the
principal objectives of the TERAPIX facility is to carry out high-level
processing (image co-addition and catalogue extraction) for the CFHTLS
survey.  The software tools described here were developed in the
context of this project\footnote{http://terapix.iap.fr/doc/doc.html},
and are described in detail in the following Sections of this paper.
These tools were designed to be as general as possible and will work on
mosaic data from other telescopes and cameras.  They are available at
the TERAPIX web site\footnote{http://terapix.iap.fr/soft}.

Our data reduction technique involves first de-trending our raw data,
then computing astrometric and photometric solutions for all input CCD
frames. These solutions are then used to re-sample and stack all CCDs
for each filter into a single contiguous mosaic, from which catalogues
are then extracted. We now describe each of these steps in turn.

\subsection{Prereductions}
\label{sec:prereductions}

All data were pre-reduced using the Fits Large Image Processing
Software ({\sc FLIPS})
package\footnote{http://www.cfht.hawaii.edu/~jcc/Flips/flips.html}.
Pre-reductions followed the normal steps of overscan, bias and dark
subtraction. Dark current for the CFH12K is negligible, around 3
$e^{-}$pix$^{-1}$~hr$^{-1}$.  Following this, all images are then
divided by a twilight flat. For data taken in $R$ and $I$ filters,
fringe removal is accomplished by subtracting a $2\sigma$ clipped
median super-flat. These superflats were typically constructed from
around $60$ science images. At this stage all images are scaled so they
have the same sky background. After pre-reductions, the gradient of the
background variation is measured across the CCD frames. This is always
less than $2\%$ and is typically $\sim 0.5\%$.

\subsection{Computing the astrometric solution}
\label{sec:astrometry}

Astrometry and relative photometry were computed using the
``astrometrix'' and ``photometrix'' tools, part of the ``WIFIX''
package for the reduction of wide--field
images\footnote{http://www.na.astro.it/$\sim$radovich/wifix.htm},
developed in the context of the TERAPIX project. In the following
sections we describe the algorithms underlying these tasks and how they
were applied to our data. An important aspect of these procedures is
that they are able to operate in an unattended mode; given the large
volume of data involved, manual interactions must be kept to a minimum.

In the following discussion, the approximate area defined by one of our
CFH12K frames is described as a 'pointing'; in the F02 field there are
four pointings in total. To reach a specified depth in a given
pointing, several images are required, and these exposures are offset
from each other in a ``dither sequence'' which contains between five
and ten exposures.

To co-add all these separate CCD images and produce a seamless output
image requires a detailed knowledge of the astrometric solution of each
CCD frame. However, computing an astrometric solution constrained
\textit{only} by an external astrometric catalogue will not produce a
sufficiently accurate solution.  This is because the intrinsic accuracy
of even the best external catalogues have rms uncertainties of around
$\sim 0.2\arcsec$ per co-ordinate. This level of accuracy (combined
with low object sky densities) make astrometric solutions derived from
these catalogues alone unsuitable for co-addition of frames. In our
survey we wish to produce a single contiguous mosaic comprising four
separate pointings in four separate bands. Overlapping images from
different filters and adjacent pointings must be registered to
sub-pixel accuracy (or an r.m.s. smaller than $\sim 0.1\arcsec$. For
these reasons, we have developed a procedure in which the astrometric
solution is constrained \textit{simultaneously} by the external
astrometric catalogue and an internal catalogue generated from matched
objects in overlapping CCDs, both from objects in the same pointing and
in adjacent pointings. Since our $I-$ band data generally have the best
seeing, and the VIRMOS spectroscopic sample will be $I-$ selected , we
compute an astrometric solution from the $I-$ band data first. We then
use this stacked image to derive solutions for all other bands.

The astrometric catalogue we used is the United States Naval
Observatory (USNO)-A2.0 \citep{1998AAS...19312003M} which provides the
position of $\sim 0.5\times10^8$ sources.  After the removal of
saturated and extended objects between 30 and 50 sources per CFH12K CCD
frame are left (the distribution of astrometric sources is not
homogeneous on the sky).  \citet{1999AJ....118.1882D}, by carrying out
comparisons with a sample of $\sim 300$ radio sources, estimate a
radial positional accuracy of the USNO-A2.0 of $\sim 0.25''$ (defined
as the radius enclosing $68\%$ of the objects).

We begin our astrometric procedure by extracting input catalogues from
the ($I$-band data) using {\sc SExtractor} \cite{1996A&AS..117..393B} on each
input (flat-fielded and bias subtracted) image.  Saturated sources were
removed. A cross-correlation procedure was used to compute the initial
offsets (generated by the dither sequence) between these catalogues and
the astrometric reference catalogue.

The astrometric solution describes the transformation between pixel
projection plane co-ordinates $(x,y)$ and spherical co-ordinates on the
sky  $\xi, \eta$, namely,  

\begin {eqnarray} 
\xi = \sum^p_{i=0} \sum^i_{j=0} A_{ij} x^{i-j} y^j  \label {eq:astsol}\\ 
\eta = \sum^p_{i=0} \sum^i_{j=0} B_{ij} x^{i-j} y^j. \nonumber 
\end {eqnarray} 
This is equivalent to the TAN projection in \citet{2002A&A...395.1077C}
but without the $\left(x^2 + y^2\right)^{0.5}$ terms.

This solution is computed as follows: firstly, for each observing run
we select an exposure for which the {\em relative} offsets between
CCDs, rotation angles and scaling factors are computed. These values
are then applied to all other exposures. In this way it is possible to
use all the available data to compute the offset to the USNO.  In order
to further minimise the possibility of errors, we separate the
exposures into groups where fields overlap by around 50\% and then
compute the offsets between exposures in the same group.  The offset to
the USNO must be the same for a given group and we take the median of
these values.
  
Next, for each CCD we match image sources with the astrometric
catalogue and recompute the linear part of the world coordinate system
(WCS). We then compute a first-guess astrometric solution
(Eq.~\ref{eq:astsol} with $p = 2$). At this stage, a database of
overlapping sources from all CCDs and pointings (``master'' catalogues)
is also constructed.

In the final step we re-compute the astrometric solution for all CCDs,
but this time including points both from the astrometric catalogues and
from the overlapping sources catalogue. The equation is solved using as
weights the positional uncertainties on each source: we assume
$0.3\arcsec$ for the USNO sources, whereas for the overlapping sources
we use the positional uncertainties given by {\sc SExtractor}.  The system of
equations is solved by iteration. Three iterations with $p=3$ produces
RMS per-coordinate uncertainties of $\sim 0.3\arcsec$ for the USNO
sources and $\sim 0.05\arcsec$ for overlapping sources.  Unfortunately,
since the offsets between exposures in the same dithering sequence are
small ($\le 20-30\arcsec$), the astrometric solution is poorly
constrained in some regions as overlapping sources are actually
extracted from the same part of the CCD and are affected by the same
distortion (discussed at length later in this section). We therefore
decided to reduce the weighting of overlapping sources from exposures
which come from the same dither sequence.  For these objects, weights
are divided by the number of frames in that dither.

\begin{figure}
\resizebox{\hsize}{!}{\includegraphics{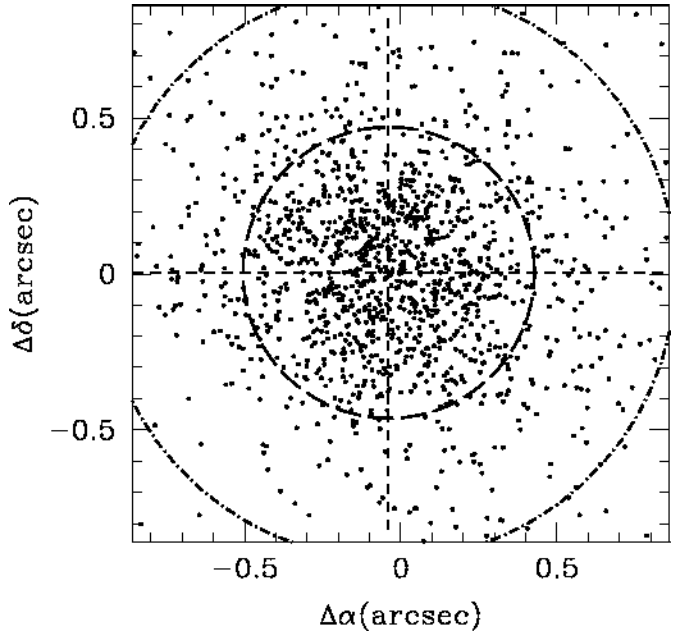}}
  \caption{Positions of USNO sources compared to
    point-like sources with $17.5<I_{AB}<22.0$ in the final
    $1.2$~deg$^2$ stacked image. The inner and outer circles (at
    $0.5\arcsec$ and $1.0\arcsec$) enclose 68\% and 90\% of all objects
    respectively. The dotted lines cross each other at the position of
    the centroid, which is at
    $(\Delta\alpha,\Delta\delta)=(-0.03\arcsec,0.00\arcsec)$.}
\label{fig:rms_usno1}
 \end{figure}

\begin{figure}
\resizebox{\hsize}{!}{\includegraphics[angle=-90]{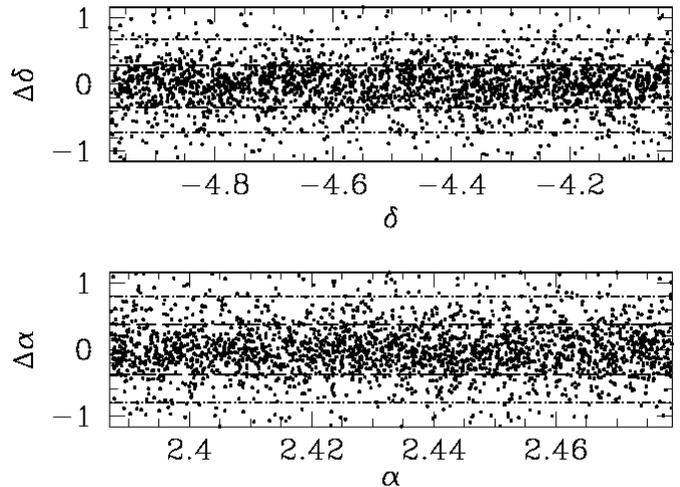}}
  \caption{Positions of USNO sources compared to point-like objects in the
    $I$ stack as a function of right ascension and declination,
    displaying the uncertainty per co-ordinate. The inner dashed lines
    enclose $68\%$ of all sources and are at
    $\vert\Delta\alpha\vert\sim0.38\arcsec$ and
    $\vert\Delta\delta\vert\sim0.32\arcsec$; the outer dot-dashed lines
    enclosing $90\%$ of all sources are at
    $\vert\Delta\alpha\vert\sim0.80\arcsec$ and
    $\vert\Delta\delta\vert\sim0.70\arcsec$.}
\label{fig:rms_usno2}
\end{figure}

How do the positions of objects in the final $I-$ stack (produced after
resampling and co-adding of all the input CCDs using this astrometric
solution) compare to our input USNO-A2 catalogue?  In
Figure~\ref{fig:rms_usno1} we plot the difference between the positions
of non-saturated, point-like sources in the $I-$ stack and their
counterparts in the USNO-A2. The inner and outer circles enclose $68\%$
and $90\%$ of all the objects respectively, and have radii of
$0.5\arcsec$ and $1.0\arcsec$. Figure~\ref{fig:rms_usno2} plots these
residuals as a function of right ascension and declination. We note
that our use of overlapping sources ensures that the accuracy of our
\textit{internal} astrometric solution is much higher, which is
necessary if we are to successfully produce coadded images, as we will
discuss now.

\begin{figure}
\resizebox{\hsize}{!}{\includegraphics{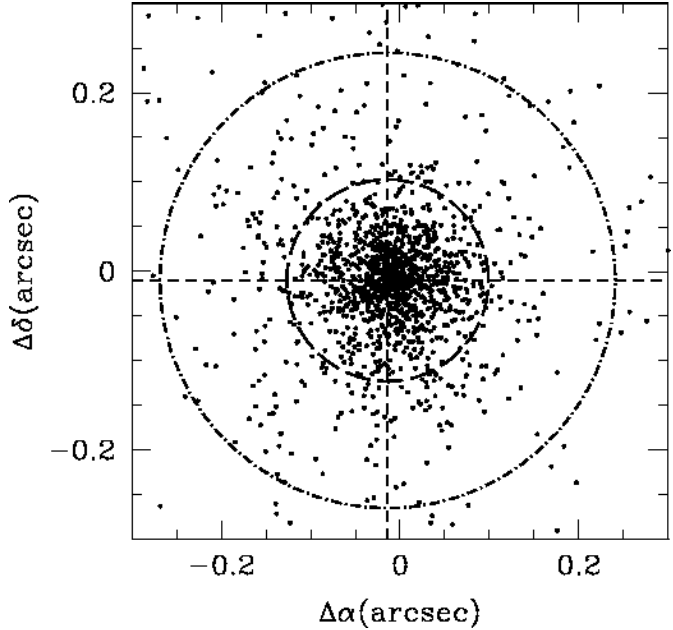}}
  \caption{Positions of unsaturated point-like sources in the $I$ stack
    image compared with objects extracted from the $B$ stacked image.
    For clarity only one point in ten is shown. The circles at
    $0.11\arcsec$ and $0.26\arcsec$ enclose $68\%$ and $90\%$ of all
    sources respectively. The dotted lines cross each other at the
    position of the centroid, which is at
    $(\Delta\alpha,\Delta\delta)=(-0.01\arcsec,-0.01\arcsec)$.}
\label{fig:rms_IVsB1}
\end{figure}

\begin{figure}
\resizebox{\hsize}{!}{\includegraphics[angle=-90]{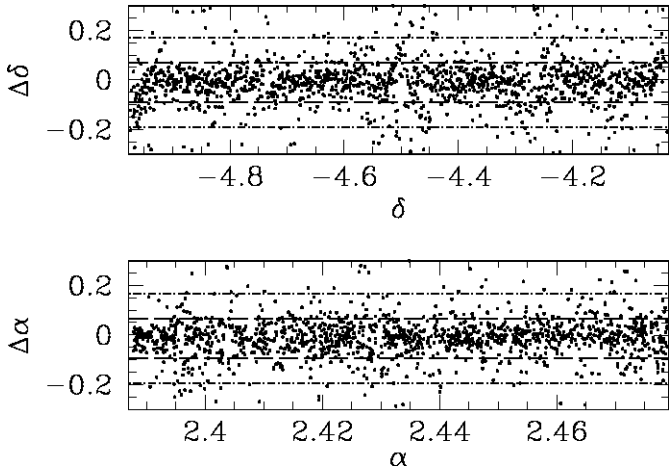}}
  \caption{Residuals between $I$ and $B$ stacks as a function of
    right ascension and declination. The inner dashed lines enclose
    $68\%$ of all sources and are at
    $\vert\Delta\alpha\vert\sim0.08\arcsec$ and
    $\vert\Delta\delta\vert\sim0.08\arcsec$; the outer dot-dashed lines
    enclosing $90\%$ of all sources are at
    $\vert\Delta\alpha\vert\sim0.18\arcsec$ and
    $\vert\Delta\delta\vert\sim0.18\arcsec$.}
\label{fig:rms_IvsB2}
\end{figure}

To compute the astrometric solution for the $BVR$ data, we use as our
reference catalogue a list of non-saturated point-like sources extracted
from the $I-$ band stack. The procedure described above is then
repeated using this catalogue. The surface density of these sources is
much higher than that of the USNO-A2, and this approach allows images
with different filters to be registered to a much higher accuracy. In
Figure~\ref{fig:rms_IVsB1} we show the positional uncertainties between
objects extracted in the $B$ and $I$ $1.2~\deg^2$ stacks. The inner
circle and outer circles plotted at $0.11\arcsec$ and $0.26\arcsec$
enclose $68\%$ and $90\%$ of all objects. These figures indicate we
have reached our required level of sub-pixel accuracy. In
Figure~\ref{fig:rms_IvsB2} these positional uncertainties are plotted
as a function of right ascension and declination.

Examining Figure~\ref{fig:rms_IvsB2}, however, a potential problem
becomes apparent. It is clear that there are regions in the stacked
images where the rms astrometric uncertainty is much higher the
average, although the area affected by these problems is small, less
than $1\%$ of the total area. Two effects are responsible for these
difficulties. The first is that the surface density of astrometric
reference sources on the sky is not homogeneous: a situation can easily
arise for example, in which there are few reference sources on one half
of a CCD. Secondly, the dither offsets we use are too small to
adequately sample the astrometric solution of the detector, and in many
cases overlapping sources come from the same regions of the same CCDs.
The net result is that in certain areas (mostly at the edges of CCD
frames) the astrometric residuals are large. However, these regions are
easily identified by comparing stacked images from each filter, and
these areas are masked. The amount of area lost from these problems is
small, less than a few percent of the total.

\subsection{Photometric calibration}
\label{sec:photometry}

The photometric calibration is carried out in two steps. In the first,
we determine an absolute calibration based on observations of standard
star fields. In the second, we combine the absolute photometric
calibration with a relative calibration computed using catalogues of
overlapping sources to derive scaling factors which will be applied
during stacking.  We now describe both these steps in turn.

The absolute photometric calibration was performed by comparisons to a
set of Landolt fields \citep{1992AJ....104..340L}. We were able to
identify at least one set of photometric images from our observations,
based on local sky conditions and stability of the object fluxes. Our
observations are scaled to these photometric images using the procedure
outlined below. Note that in this work, we do not attempt to calibrate
each individual CCD frame independently, but instead calibrate the
scaled, flat-fielded images (i.e., the entire mosaic). In other words,
we assume that each CCD has the same colour equation. Since our
observations were made, the CFH12K photometric performance has been
extensively characterised as part of the CFHT queue observing program
{\sc Elixir}, which has determined zero-points and colour equations
based on observations of several hundred standard star
frames\footnote{http://www.cfht.hawaii.edu/Instruments/Elixir/filters.html}.
The assumption that each CCD has the same photometric equation produces
an overall photometric residual of less than $4\%$. These
investigations also indicate that the CFH12K $BVR$ filters have
negligible colour terms (less than 0.03 in $(B-V)$ and $(V-R)$ with
respect to the Johnson Kron-Cousins system). Only in $I$ was a
significant colour term found (quoting from the ELIXIR web pages):

\begin{equation}
I_{kc}=i_{cfh12k}+26.185-0.04(\kappa-1)+0.107(R-I), 
\end{equation}

where $\kappa$ represents the airmass, $R,I$ are Johnson magnitudes and
$I_{kc}$ is the magnitude in Johnson Kron-Cousins system.

These zero-points and colour equations are broadly consistent with
those determined from our survey data, but are of higher accuracy. Note
that we do not actually use these colour equations in this work, and
present all our magnitudes in the CFH12K instrumental system, defined
by the combination of CFH12K detectors, filters and telescope. We do
however convert our instrumental magnitudes to the ``AB'' magnitude
system, \citep{1974ApJS...27...21O} as follows:
$B_{AB}=b_{cfh12k}-0.094$, $V_{AB}=v_{cfh12k}-0.007$,
$R_{AB}=r_{cfh12k}+0.218$ and $I_{AB}=i_{cfh12k}+0.433$, computed from
the CFH12K filter curves. Our zero-points are corrected for galactic
extinction of $E(B-V)=0.027$ estimated at the centre of the field
provided by the maps of \citet{1998ApJ...500..525S}. As the galactic
extinction is quite uniform over the field, we do not attempt to apply
position-dependent magnitude corrections to our data.

After computing the absolute photometric calibration, and identifying
photometric exposures in our data, we next compute the relative scaling
exposure-to-exposure. The zero-point of each image $i$ may be written
as the sum of two components:
\begin{equation} 
Z_i = z^{\rm ph} + z^{\rm r}_i,  
\end{equation} 
where the photometric zero point $z^{\rm ph}$ is determined using
standard stars and the relative zero point $z^{\rm r}_i$ incorporates
the airmass correction and other changes in the atmospheric
transparency.  We assume that the relative zero point is the same for
all the CCDs in a mosaic and that at least one image has been taken in
photometric conditions, so that it may be used as reference. It is
computed by minimising the average difference in magnitude of
overlapping sources in different frames . Our approach is similar to
that of \cite{1998PASP..110.1464K} except that we do not derive our
coefficients by direct solution of a matrix inversion but by an
iterative minimisation process.
 
The astrometric solution detailed in Section~\ref{sec:astrometry}
provides a database of overlapping sources which can be used to compute
the photometric offsets. We first extract from these catalogues pairs
of objects from overlapping frames with magnitudes $m_i,m_j$ and
photometric uncertainties $w_i,w_j$. For each pair we compute
\begin {equation} 
\delta m_{ij} = \frac{\sum{w_i m_i}}{\sum {w_i}} -  
       \frac{\sum{w_j m_j}}{\sum {w_j}} .  
\label{eq:photoff} 
\end{equation} 

Our aim is to find the set of values $z_i$ which minimises
\begin {equation} 
\delta m^*_{ij} = <m>_i + z_i - (<m>_j + z_j) = \delta m_{ij} + (z_i - z_j) 
\end{equation} 
 
For each frame $i$ with overlapping sources in other $N_{\rm ov}$ frames, 
the zero point is: 
\begin {equation} 
 Z_i = z^{\rm ph} +  
 \frac{\sum_j^{N_{\rm ov}} {z_j -  \delta m_{ij}}}{N_{\rm ov}}  
   - \frac{\sum_{k}^{N_{\rm ph}} {z_k}}{N_{\rm ph}},   \label{eq:photsol} 
\end{equation} 
where we assume that $N_{\rm ph}$ frames are known to be photometric
and that for them the average relative zero point is null. The
effective airmass $\kappa$, of the coadded image is:
\begin{equation} 
\kappa = \frac{\sum_{k}^{N_{\rm ph}} {\kappa_k}}{N_{\rm ph}} 
\end{equation} 
 
Eq.~\ref{eq:photsol} is iterated until $\frac{\sum_i^N {(\delta
    m^*_{ij})^2}}{N}$ converges or a maximum number of iterations (50)
is reached.

%In Figure~\ref{fig:photomplot1} we illustrate the effect of this
%correction process from the ``wide'' F14-I band data. We note that most
%of the overlapping images were taken on the same night and,
%additionally, the airmass correction is small. For this reason, the
%mean value does not change before and after the application of the
%correction. The dispersion reduces from $\sim 0.3$ to $0.2$ magnitudes
%after the application of this correction.

%\begin{figure}
%\resizebox{\hsize}{!}{\includegraphics{./eps/magdiff14_v2.ps}}
%  \caption{Absolute value of the magnitude difference between objects in
%    overlapping CCDs before and after the application of the
%    corrections described in Section~\ref{sec:photometry} (dotted and
%    solid lines respectively).}
%\label{fig:photomplot1}
%\end{figure}

\subsection{Image resampling, stacking and flux scaling}
\label{sec:image-resampl-stack}

Once the astrometric and photometric solutions have been computed for
all the input CCDs, these images are combined to produce the final
stack. This is carried out in a two-step process using a new image
resampling tool, {\sc SWarp}\footnote{http://terapix.iap.fr/soft/swarp}
\citep{EB2002}.

In the first step, each input CCD is resampled and projected onto a
subsection of the output frame (which covers all input images) using
the astrometric solutions computed in Section~\ref{sec:astrometry}. We
use a tangent-plane projection, which is adequate given the relatively
small size of our field. The flux scalings discussed in the previous
Section are also applied at this stage. An interpolated, projected
weight-map is also computed which incorporates information derived from
bad pixel maps.  Both for the weight-map and image data, interpolation
is carried out using a ``Lanczos-3'' interpolation kernel. This kernel
corresponds to a sinc function multiplied by a windowing function. It
provides an optimal balance between preserving the input signal and
noise structure and limiting ringing artifacts on image discontinuities
(for instance, around bright stars and satellite trails).
Additionally, at this stage a local sky background is computed at each
pixel and subtracted, which ensures that the stacked image is free of
large-scale background gradients.

In the final stage, all resampled images are coadded (with each pixel
weighted by the appropriate weight map) to produce the final output
image. Pixels are coadded using a median, which although sub-optimal
for signal-to-noise properties, provides the best rejection of
satellite trails and other cosmetic defects. In addition to the final
coadded image, a total weight map is also produced containing
information concerning how often each pixel has been observed. This
weight-map is used during the detection and catalogue generation steps,
outlined in the following sections. An important aspect of the
resampling and stacking stage is that a single, large contiguous image
is produced which contains all data, resulting in only one catalogue
for each filter. No merging of catalogues is required. We set the size,
pixel scale and orientation of the $BVR$ data to match the one computed
for the $I$ data so that catalogues are easily extracted from these
other images.

\subsection{Catalogue preparation}
\label{sec:catal-prep}

To prepare merged $BVRI$ catalogues we locate objects using a separate
detection image and then perform photometry in each bandpass at the
positions defined by the detection image. This image is constructed
using the $\chi^2$ technique outlined in \cite{1999AJ....117...68S},
namely

\begin{equation}
 y = \sum_{i=1}^{n}(a_i/\sigma_i)^2,
\label{eq:1}
\end{equation}

where $a_i$ represents the background-subtracted pixel value in filter
$i$, $\sigma_i$ the rms noise at this pixel and $n$ is the number of
filters (four in this case).

In the absence of sources, the distribution of sky values in this
stacked image can be described as $\chi^2$ distribution with $n$
degrees of freedom. Detection and thresholding, therefore, have
relatively simple interpretation in probabilistic terms. Our primary
motivation in using this technique is to simplify the generation of
multi-band catalogues and to reduce the numbers of spurious detections.

We produce this $\chi^2$ image from the four stacked images using {\sc
  SWarp} which performs the local estimations of $\sigma_i$ and the sky
background at each pixel (note that in version 1.21. of {\sc SWarp},
used here, the output pixels in the generated $\chi^2$ images are
actually $S=\sqrt{(y/n)}$).  For the $\chi^2$ technique to work
reliably, seeing variations across input images must be small (less
than $\sim 20\%$) and moreover there must be the same number of input
images at each pixel (a consequence of this is that detections may not
be reliable in parts of the image where there is no coverage in all
four bands). We have compared magnitudes of objects recovered with and
without the use of the $\chi^2$ detection image and we find no evidence
that it introduces any magnitude bias.

For objects to be included in our catalogues, they must contain at
least three contiguous pixels with $S>2$ in the $\chi^2$ image. This
corresponds to a per-pixel detection threshold of $2.8\sigma$. With
this threshold we detect in total 460,447 sources. We chose this
somewhat conservative threshold in order to minimise the numbers of
false detections. However, as we shall see in the following sections,
our catalogues are essentially complete at $I_{AB}=24$. This threshold
is close to the 'optimal' value where the probability of misclassifying
object pixels as sky and sky pixels as objects is minimised (see
Figure~1. of \citeauthor{1999AJ....117...68S}).

In our catalogues, magnitudes are measured using
\cite{1980ApJS...43..305K}-like elliptical aperture magnitudes computed
using {\sc SExtractor} (the \texttt{magauto} parameter).  These magnitudes
provide the most reliable measurement of the object's total flux (based
on simulations carried out on our images), although they may be subject
to blending effects if there are near neighbours. This is not a concern
for us because our images in all bandpasses are far from the confusion
limit and the numbers of objects affected are very small. We set a
minimum Kron radius of $1.2''$, which means that for faint unresolved
objects (where the Kron radius can be difficult to reliably determine),
our ``total'' magnitudes will revert to simple aperture magnitudes.
Colours are always measured at the same location (defined by the
detection image) on each stack.  Because there are many large, extended
objects on our images, we also use \texttt{magauto} magnitudes to
measure object colours. Once again, tests show that these colours
revert to ``aperture'' colours for the majority of faint, unresolved
objects in our catalogues. Measurements show that for total magnitudes
in the range $18.0<I_{AB}<24.0$ the difference between the median galaxy
colour using total and aperture magnitudes is less than 0.05
magnitudes.

We use the {\sc SExtractor} \texttt{flux\_radius} parameter to identify
point-like sources, using object profiles measured in the $I-$band
images with centroids derived from the $\chi^2$ image (of course, given
that we have $BVRI$ photometry for all objects, a complementary
approach would be to identify stellar sources based on best-fitting
spectral templates; application of this technique to our data will be
described in future papers). This parameter, denoted as
$r_{1\over{2}}$, measures the radius which encloses a specified
fraction of the object's total flux (in this case, $50\%$).  Fainter
than saturation ($I_{AB}\sim 18$), point-like objects have a flux
radius independent of magnitude.  In our data, an added complication is
that the seeing differs by around $20\%$ across the mosaic, which means
this locus is slightly dispersed.  To account for this, we used an
``adaptive'' classification technique.  In this procedure, we divide
the field into many sub-areas and compute the mode of $r_{1\over{2}}$
for objects within each box. Since point-like objects all have the same
half-light radius this provides a robust measurement of the local
seeing value. A search is then carried out for the bin which has a
value of less than $20\%$ of the modal bin; all objects with
$r_{1\over{2}}$ smaller than this are classified as stars.

\begin{figure}
  \resizebox{\hsize}{!}{\includegraphics{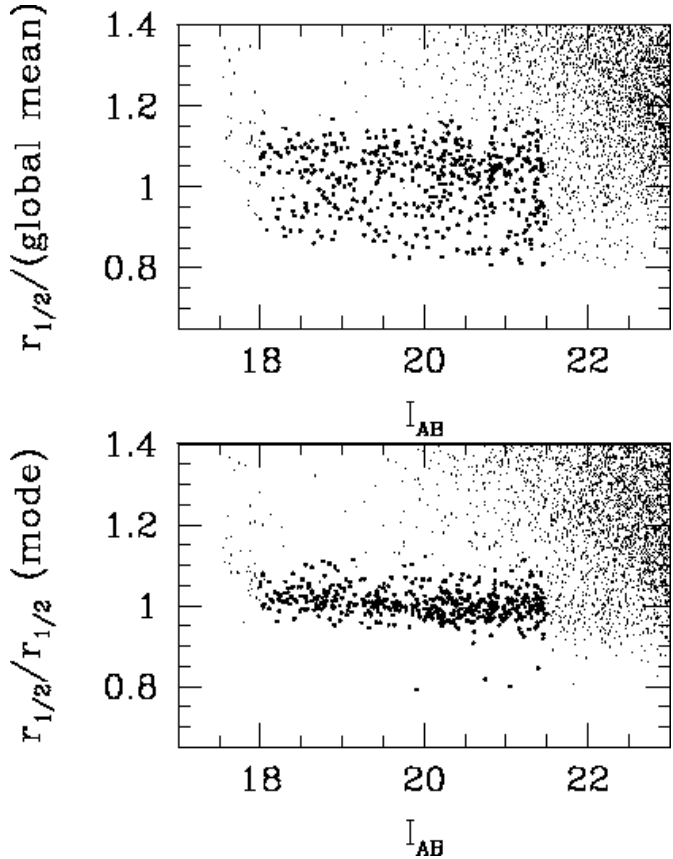}}
  \caption{Object half-light radius as a function of $I_{AB}$ magnitude. 
    In the upper panel, we show the half-light radius in pixels
    (normalised by full-field mean of 2.5 pixels) over the full mosaic
    both for point-like (solid points) and extended sources (light
    points).  In the lower panel each object's half-light radius has
    been normalised by the local mode, computed using the adaptive
    classifier (in both panels, for clarity, only a fraction of all
    objects are shown).}
\label{fig:r2norm}
\end{figure}

Figure~\ref{fig:r2norm} demonstrates how this method works in practice.
In the upper panel we show the half-light radii values as a function of
$I_{AB}$ magnitude for all objects, normalised by the approximate mean
value for the full field. In the lower panel, each object's half-light
radii has been divided by the local measurement. The heavy dots in both
panels indicate the point-like objects located using the classifier. The
separation between resolved and unresolved sources is considerably
improved, and distinguishing between the two seems feasible until
$I_{AB}\sim22$, although in this work we adopt a conservative limit of
$I_{AB}\sim 21.5$. Beyond this limit no classification is attempted.

\begin{figure}
\resizebox{\hsize}{!}{\includegraphics{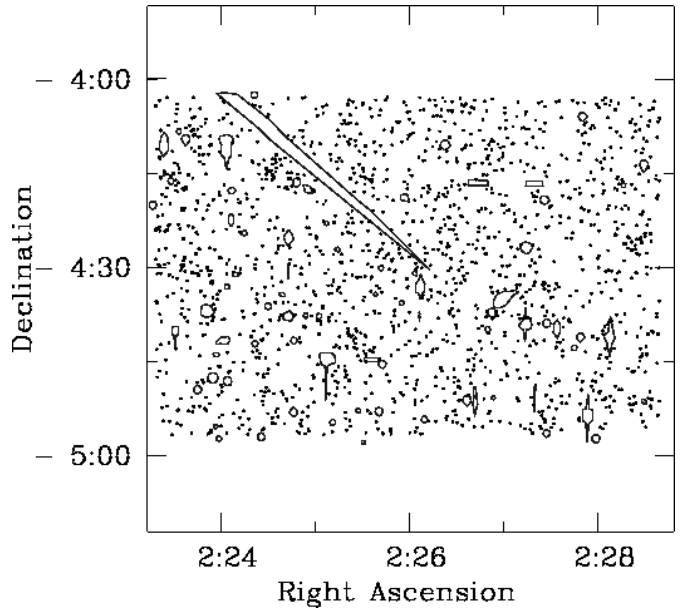}}
  \caption{Distribution of bright galaxies. The points show galaxies
    (extended sources) with $18.0<I_{AB}<21.0$. Masked regions are also
    indicated.}
\label{fig:f02distrib}
\end{figure}

Finally, the catalogue was visually inspected by over-plotting on the
images all non-saturated objects with $I_{AB}<24.0$.  Areas
contaminated by residual fringing, satellite trails and scattered light
were flagged by drawing polygonal masks around them. All large bright
stars and their corresponding diffraction spikes were also masked. In
Figure~\ref{fig:f02distrib} we show the distribution of galaxies with
$18.0<I_{AB}<21.0$. Masked regions, outlined by polygons, are also
indicated. In the magnitude range $18.0<I_{AB}<24.0$ we detect 90,951
extended objects not in the masked regions or at the borders of the
image. In this area and magnitude range there are 2857 point-like
sources. The total area of our survey, after masking is $1.18$~deg$^2$.

\section{Data quality assessment}
\label{sec:data-qual-assessm}

In these sections we describe and present a series of quality
assessment tests carried out on the $BVRI$ catalogues prepared in the
previous sections. 

\subsection{Surface brightness selection effects and catalogue limiting
  magnitudes}
\label{sec:limit-magn-catal}

Our objective here is to determine the photometric completeness of our
survey in the plane defined by surface brightness and total magnitude.
The importance of surface brightness selection effects in photometric
surveys has already been discussed by several authors
\citep{1995ApJ...455...50L,1993ApJ...403..552Y}.  In this work, an
important objective of our simulations is to ensure that our data is
fully complete and free from surface brightness selection effects at
$I_{AB}=24.0$, the spectroscopic limit of the VLT-VIRMOS deep survey.

To estimate how surface brightness selection effects could affect our
catalogues a variety of approaches may be used. For example, we might
try to produce simulated objects with a variety of light profiles, such
as exponential disks or $r^{1/4}$-law bulges. These objects could then
be dimmed based on some physically motivated picture of how galaxies
evolve and then coadded back into our image frames and their object
parameters re-measured.  In what follows, however, we adopt a purely
empirical approach. We start by extracting $BVRI$ sub-areas of 100
bright, isolated galaxies from a $1024\times1024$ subsection of the
field. Our objective is to quantify, at each combination of magnitude
and total surface brightness, how many objects are recovered by our
selection procedure.  Some of the surface-brightness/magnitude
combinations are not physical, but we wished to introduce as few
assumptions as possible about the evolution of the faint galaxy
population into our simulations. A consequence of this approach, of
course, is that interpreting the overall completeness requires
additional information concerning where the faint galaxy population
resides in the surface-brightness/magnitude plane.

To relocate our galaxies in the surface brightness/magnitude plane we
reduce their total magnitude and surface brightnesses by dividing each
object by some number to scale it to a fainter total magnitude. To
reduce the surface brightness of the galaxies while keeping the same
total magnitude, the galaxies were stretched using bilinear
interpolation.  After these alterations, the galaxies are added back
into the subsections extracted from each of four bandpasses. Following
this, objects are detected using the same procedures as in the real
catalogues. We generate a $\chi^2$ detection sub-image and search for
objects using our detection threshold (corresponding to three connected
pixels having $S>2$). For each combination of $I_{AB}$ total magnitude
and peak surface brightness (defined as the surface brightness of the
brightest pixel in the recovered object) we measure the fraction of
galaxies retrieved.  To evaluate the numbers of spurious detections,
all the images are multiplied by $-1$. As before, the $\chi^2$ image is
generated and the flux measurement is carried out using {\sc
  SExtractor} in dual image mode on the negative images. The number of
initial detections will be the same since the chi-squared detection
image generated from the negative images is the same as the chi-squared
image generated from the positive images. However, since the
measurement image is now negative, positive fluxes will be measured
only for the false detections.

\begin{figure}
\resizebox{\hsize}{!}{\includegraphics{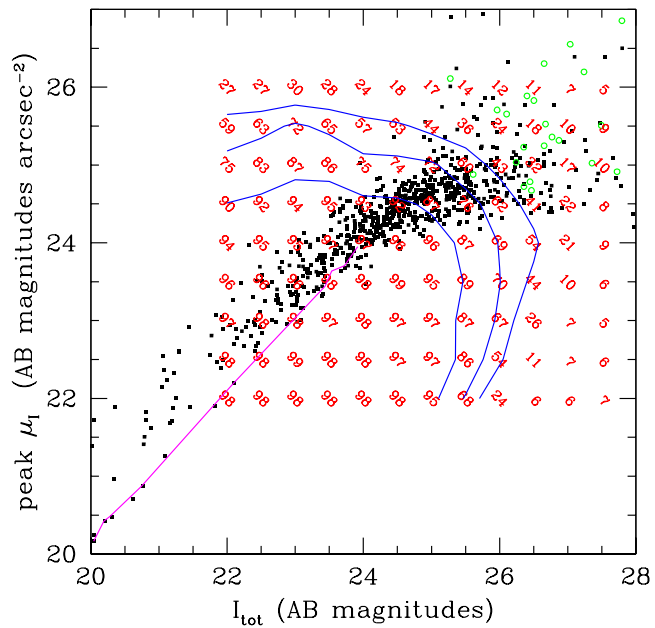}}
  \caption{Peak surface brightness as a function of  $I_{AB}$ total
    magnitude. Detected galaxies are shown as the black dots, the open
    circles show the spurious detections. The stellar locus is
    represented by the solid line. Most galaxies lie nearby to this
    locus. The contour lines indicate the $50\%$, $70\%$ and $90\%$
    completeness limits. The numbers represent the fraction of objects
    recovered at each combination of total magnitude and peak surface
    brightness.}
\label{fig:limits}
\end{figure}

The results of these simulations are illustrated in
Figure~\ref{fig:limits} where they are plotted in the plane defined by
total magnitude and peak surface brightness. Filled dots show the
position of all objects in the field; the solid line shows the location
of the stellar locus. For each combination of total magnitude and peak
surface brightness, the numbers indicate the percentage of objects
which are recovered. The contour lines indicate the $50\%$, $70\%$ and
$90\%$ completeness limits. The open circles show the spurious
detections; we note that these spurious detections only become
significant faintwards of $I_{AB}\sim25$.

For point-like stellar objects at $I_{AB}=24.0$ our survey is essentially
complete, with approximately $90\%$ of objects recovered at this
combination of magnitude and surface brightness. At this magnitude our
recovery fraction drops to $50\%$ for objects with peak surface
brightnesses of around $\sim 25.5$~mag~arcsec${^2}$.
\citet{1995ApJ...455...50L} have computed the tracks of typical
galaxies in the surface-brightness magnitude plane, and their Figure~7
shows that most ``normal'' galaxies would be easily detected by our
survey. Objects with extremely large half-light radii, such as the
prototypical ``Malin 1'' low surface brightness galaxy could fall below
our detection limits at higher redshifts.  However, space-based
observations suggests that the numbers of such extended, low-surface
brightness objects at these magnitudes is actually quite low
\citep{1999ApJ...519..563S,1998AJ....115.1400F}.  Most of the objects
we detect in the real data (filled points in Figure~\ref{fig:limits})
fall in quite a narrow locus in the surface-brightness/magnitude plane.
Moreover, if we compare our total galaxy number counts at $I_{AB}=24.0$
with an average of measurements in North and South Hubble Deep Fields
\citep{2001MNRAS.323..795M} (presented in
Section~\ref{sec:galaxycounts}) we see that we are in good agreement at
$I_{AB}=24$ with these much deeper counts. Based on these
considerations, we conclude that our survey is essentially free of
selection effects until at least $I_{AB}\sim 25$. Finally, we note that
our simulations also show that at this limit our survey is free from
spurious detections (represented as the open circles in
Figure~\ref{fig:limits}).

We have also carried out a set of simulations based on adding
progressively fainter simulated stellar sources to a sub-section of the
image stacks. The magnitudes at which we recover $50\%$ of the input
sources is $26.5$,$26.2$,$25.9$ and $25.0$ for $BVRI$ filters
respectively.

\subsection{Catalogue completeness as a function of position}
\label{sec:poscomp}
An important issue to address is how homogeneous the detection and
measurement limits are over the entire mosaic, and how stable the images'
noise properties are as a function of location. To estimate this, we
carried out a series of simulations aimed at measuring the
incompleteness as a function of position for the full $I-$ band mosaic.
In these simulations, we extracted $\sim 200$ sub-areas arranged in a
grid over the mosaic. For each of these sub-areas, we added artificial
stars at random positions using as a local seeing value the
measurements derived in Section~\ref{sec:catal-prep} (our intention in
this experiment is not to provide a realistic assessment of the
absolute value of the incompleteness but rather its positional
variation, which is why we adopt the simplifying assumption of stellar
profiles). Detection and measurement was then carried out using
{\sc SExtractor}, using detections weighted by the local detection
weight-map. For several slices in magnitude, we measured what fraction
of stars we recover at each position.

\begin{figure}
\resizebox{\hsize}{!}{\includegraphics{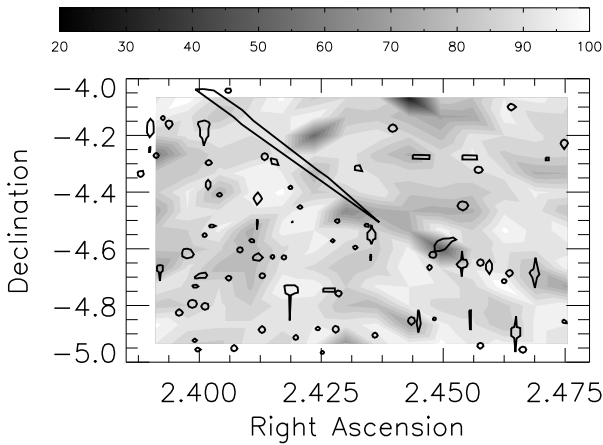}}
  \caption{Measured incompleteness as a function of position (decimal
    declination and right ascension) for the $I$ band mosaic. The
    grey-scale contours indicate, at each position, the fraction of
    stellar objects which were recovered.  Input magnitudes were
    distributed uniformly in the range $23.75<I_{AB}<24.25$. The thick
    lines show the location of the masked regions.}
\label{fig:compmap}
\end{figure}

Our results are illustrated in Figure~\ref{fig:compmap}, where we plot
the recovery fraction as a function of position for input objects in
the magnitude range $23.75<I_{AB}<24.25$. Over most of the field, the
recovery fraction is $> 90\%$ (in good agreement with the result
derived for point-like sources in the previous section).  There are
small regions where the completeness can be significantly lower; upon
investigation, these areas appear to be populated by bright stars which
were not excluded from the simulation. From these tests we conclude
that there is no significant variation in completeness over the mosaic
at $I_{AB}=24.0$.

\subsection{Galaxy and star counts}
\label{sec:galaxycounts}

In this Section we compare the source counts of our catalogues of
galaxies and stars (in reality, point-like and extended sources) to
published compilations. The filled squares in Figure~\ref{fig:icounts}
shows $I_{AB}$-selected stellar number counts compared to the
predictions of the model of \cite{AR}, computed at the galactic
latitude of our survey. In the magnitude range $18<I_{AB}<21.5$ we
measure the slope of the stellar number counts, $d\log N/dm$, as $\sim
0.14$. Our stellar counts and slope measurement are in in excellent
agreement with model predictions (a detailed discussion of the
properties of the stars in the survey will be presented in a future
paper).

\begin{figure}
\resizebox{\hsize}{!}{\includegraphics{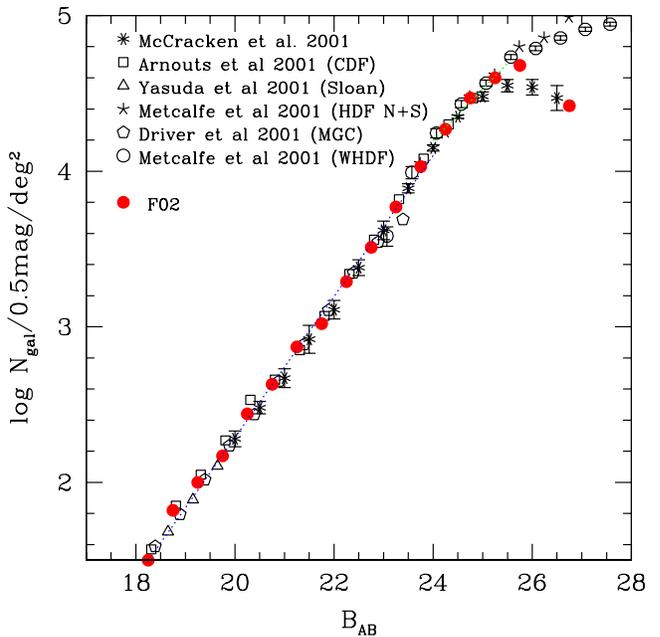}}
  \caption{$B_{AB}$-selected galaxy counts (filled
    symbols), compared to a selection of recent results from the
    literature. The dotted lines shows the best fitting slope computed
    in the magnitude ranges $20.0<B_{AB}<24.0$ and $24.0<B_{AB}<25.5$. Error
    bars are smaller than our symbols and are not shown.}
\label{fig:bcounts}
\end{figure}

\begin{figure}
\resizebox{\hsize}{!}{\includegraphics{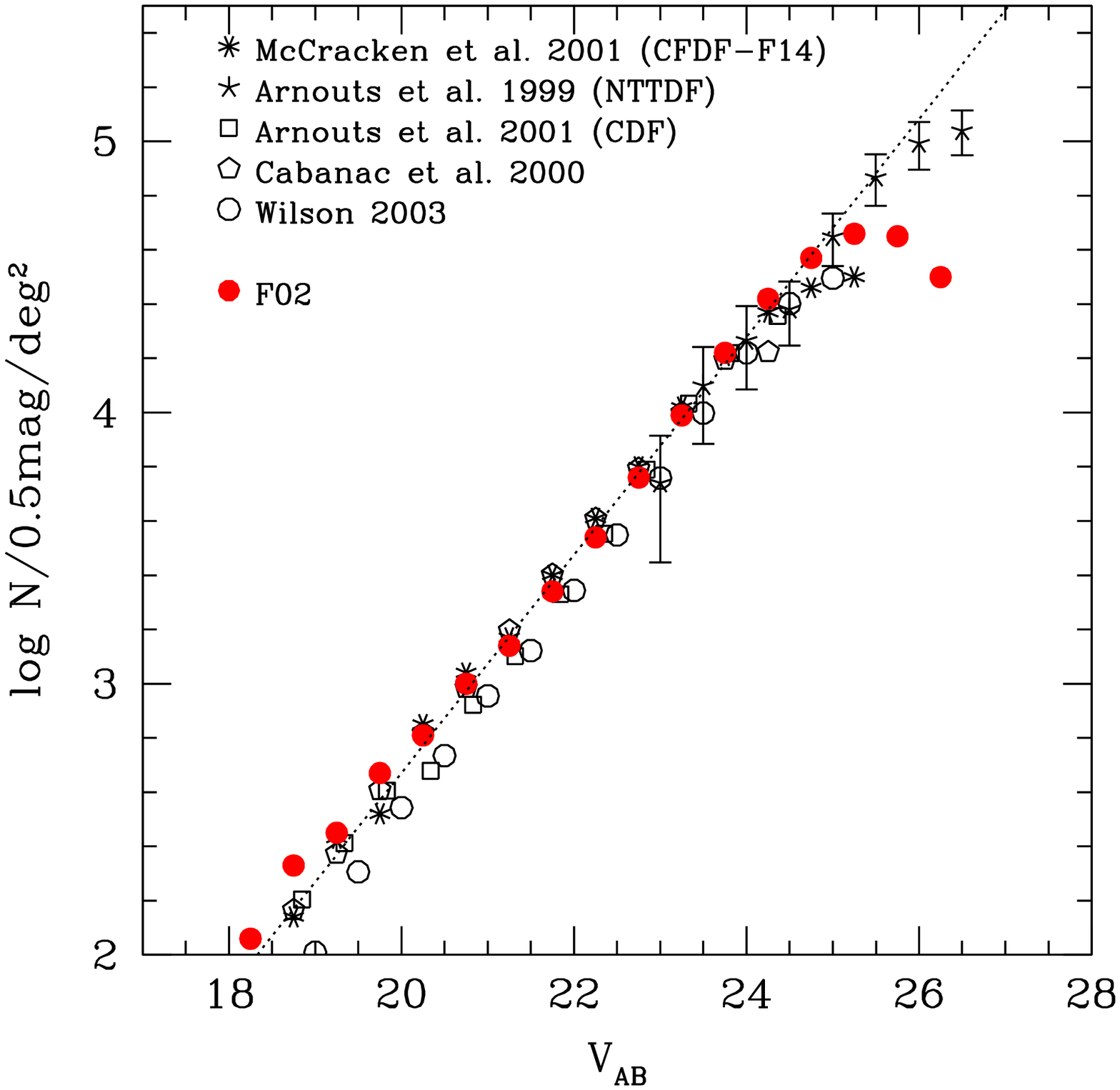}}
  \caption{$V_{AB}$-selected galaxy counts (filled
    symbols), compared to a selection of recent results from the
    literature. Error bars are smaller than our symbols and are not
    shown.}
\label{fig:vcounts}
\end{figure}

\begin{figure}
\resizebox{\hsize}{!}{\includegraphics{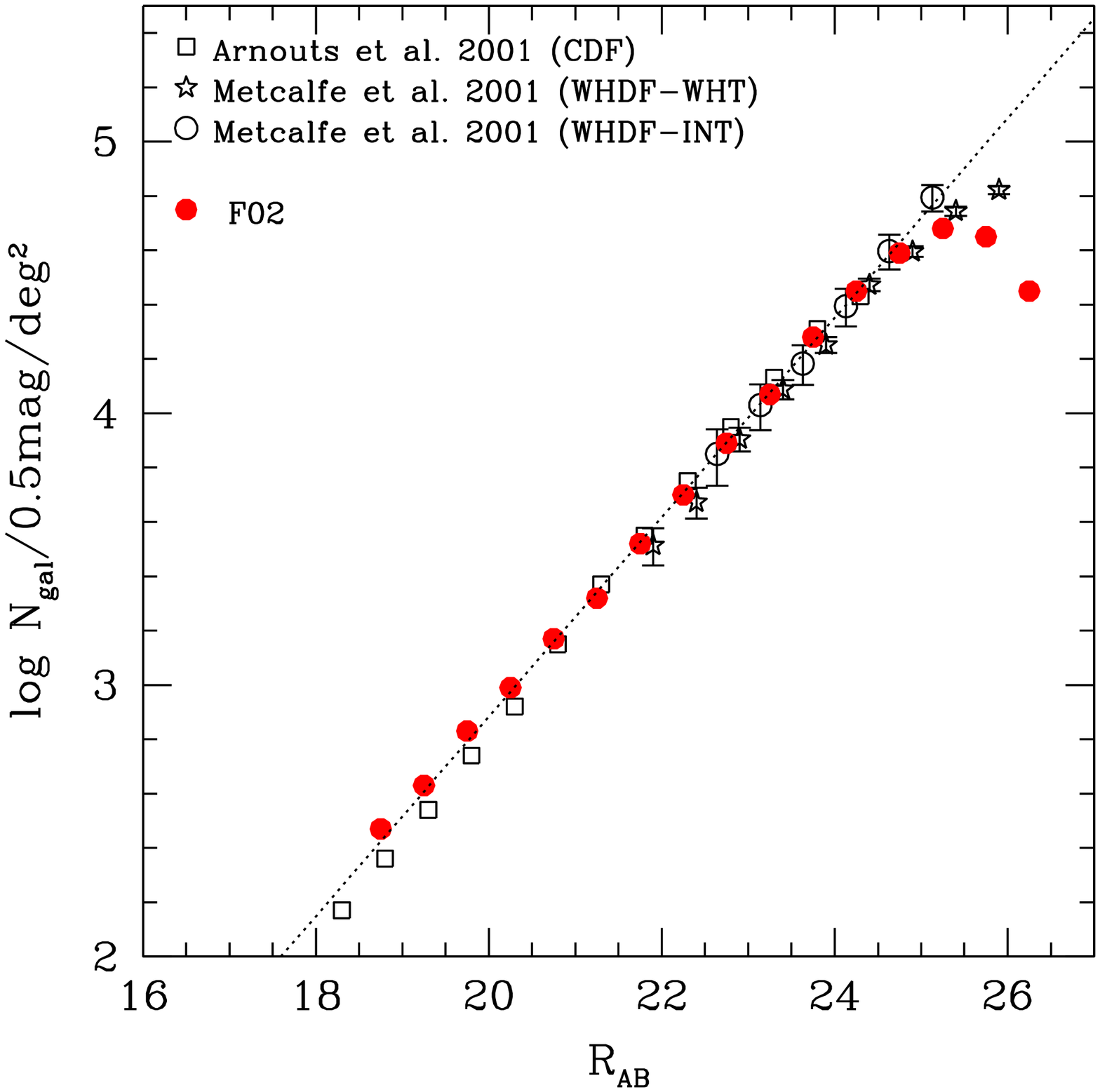}}
  \caption{$R_{AB}$-selected galaxy counts (filled
    symbols), compared to a selection of recent results from the
    literature. Error bars are smaller than our symbols and are not
    shown.}
\label{fig:rcounts}
\end{figure}

\begin{figure}
\resizebox{\hsize}{!}{\includegraphics{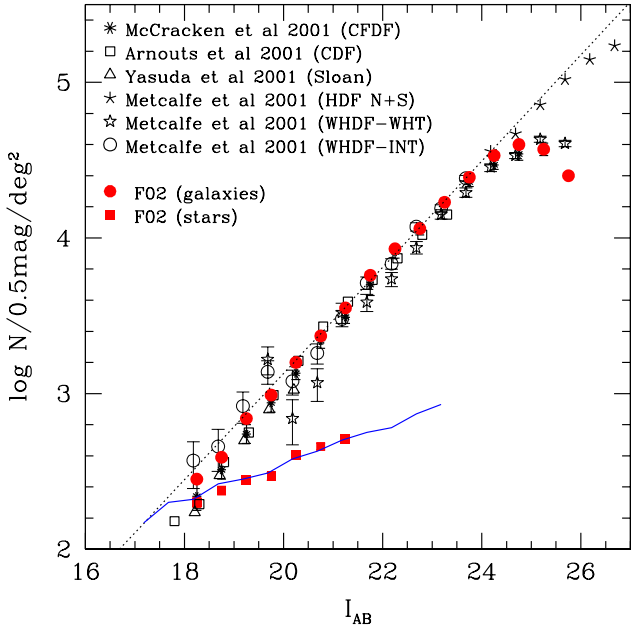}}
  \caption{$I_{AB}-$ selected galaxy and star counts (filled
    circles and squares respectively), compared to a selection of
    recent results from the literature (open symbols). For the galaxy
    counts, in all cases our (Poisson) error bars are smaller than the
    symbols and are not plotted. The dotted line shows the best fitting
    slope computed in the magnitude range $20<I_{AB}<24$. The solid
    line shows the predictions of the galactic model of Robin et al.,
    computed at the galactic latitude of our survey. }
\label{fig:icounts}
\end{figure}

Differential galaxy number counts per square degree per half magnitude
for $B$, $V$, $R$, and $I$ filters derived from our catalogue are
represented as the filled circles in Figures~\ref{fig:bcounts},
\ref{fig:vcounts}, \ref{fig:rcounts} and $\ref{fig:icounts}$. These
counts are also listed in Table~\ref{tab:f02.counts}. We do not correct
our counts for incompleteness, and show ``raw'' measurements.  For our
literature comparison, we include only measurements from recent digital
surveys and omit earlier photographic works.  We note how well the
galaxy count measurements now agree well between the different surveys,
in particular for the $B-$band measurements. It appears that the large
scatter between different groups in previously presented compilations
arose from uncertainties in photographic magnitude scales at the bright
end and from the small fields of view of the previous generation of
deep CCD surveys at the faint end. By contrast, in many large surveys
today, systematic errors arising from absolute zero-point uncertainties
dominate in almost every magnitude bin.

In general, our agreement with the literature measurements is
excellent. In the $V-$ band, our counts are in good agreement with
those of \cite{2000astro.ph..7184C}, the CFDF measurements and those of
\cite{2001A&A...379..740A}. The counts from \cite{2003ApJ...585..191W}
appear to be systematically below all other measurements.

We have also carried out a more detailed bin-by-bin
comparison with the $I_{AB}$ galaxy counts determined in the CFDF
survey, which has comparable total area ($\sim 1$~$\deg^2$) to the
present work. The CFDF counts differ at most from the present work by
$\sim 10\%$ per bin. Given the large numbers of galaxies in each bin,
this error is almost always larger than the combination of $\sqrt N$
Poisson counting errors and fluctuations induced by large scale
structure (which is small for the fainter bins). Assuming this
difference arises from uncertainties in the absolute zero-points of the
CFDF and the present work, this would correspond to absolute zero-point
uncertainty of $\sim0.05$ magnitudes (estimated simply by calculating
what zero-point shift minimises the difference in counts between the
two surveys).  This is in agreement with what we have estimated in
Section~\ref{sec:photometry} from observations of Landolt standard star
fields.

We have also determined the slope $\alpha$ ($d\log N/dm$) of the galaxy
counts, and these are reported in Table~\ref{tab:f02:slopes}. Our $B$
and $I$ slopes agree well with reported CFDF survey values of $\sim
0.35$ and $\sim 0.47$ respectively.  Fainter than $B_{AB}\sim24$ we
observe a slope flattening in the $B-$: for $24<B_{AB}<25.5$, we find
$d\log N/dm \sim 0.3$, similar to that found by \cite{MSFR}.

\begin{table*}
\begin{tabular}{*{10}{c}}
\\& \multicolumn{2}{c}{\bf{$B$}} & \multicolumn{2}{c}{\bf{$V$}}&
\multicolumn{2}{c}{\bf{$R$}} & \multicolumn{2}{c}{\bf{$I$}} \\
\\
{AB Mag} & {$N_{gal}$} & $\log N_{gal}$~$\deg^{-2}$ & $N_{gal}$ &
$\log N_{gal}\deg^{-2}$& {$N_{gal}$} & $\log N_{gal}$~$\deg^{-2}$ & $N_{gal}$ &$\log N_{gal}\deg^{-2}$\\
\\
 18.0-18.5  & 37  &  1.50  &     137 & 2.06 &   303 & 2.41 & 332 &  2.45 \\
 18.5-19.0  & 78  &  1.82  &     251 & 2.33 &   346 & 2.47&  462 &  2.59 \\
 19.0-19.5  & 119 &  2.00  &     335 & 2.45 &   508 & 2.63&  811 &  2.84 \\
 19.5-20.0  & 176 &  2.17 &      550 & 2.67 &  793 & 2.83& 1156 &  2.99 \\
 20.0-20.5  & 328 &  2.44  &     767 & 2.81 &  1157 & 2.99& 1890 &  3.20 \\
 20.5-21.0  & 508 &  2.63  &    1182 & 3.00 &  1732 & 3.17& 2796 &  3.37 \\
 21.0-21.5  & 882 &  2.87  &    1625 & 3.14 &  2476 & 3.32& 4168 &  3.55 \\
 21.5-22.0  & 1247 &  3.02 &    2565 & 3.34 &  3930 & 3.52& 6759 &  3.76 \\
 22.0-22.5  & 2312 &  3.29 &    4137 & 3.54 &  5922 & 3.70& 9992 &  3.93 \\
 22.5-23.0  & 3813 &  3.51 &    6831 & 3.76 &  9241 & 3.89& 13676 &  4.06\\
 23.0-23.5  & 7014 &  3.77 &   11450 & 3.99 &  14006& 4.07& 19863 &  4.23\\
 23.5-24.0  & 12771 & 4.03 &   19609 & 4.22 &  22235& 4.28& 28828 &  4.39\\
 24.0-24.5  & 22130 & 4.27 &   30850 & 4.42 &  33632& 4.45& 39874 &  4.53\\
 24.5-25.0  & 34448 & 4.47 &   43593 & 4.57 &  46362& 4.59& 47079 &  4.60\\
 25.0-25.5  & 47255 & 4.60 &   54457 & 4.66 &  56686& 4.68& 43459 &  4.57\\
\end{tabular}                               
\caption{Differential galaxy number counts. Counts are
  shown in half-magnitude bins selected in $(BVRI)_{AB}$. Also shown are the logarithmic counts, normalised to the
  effective area of the survey field ($1.18$~deg$^2$).}
\label{tab:f02.counts}
\end{table*}

\begin{table}
\begin{tabular}{*{3}{c}}

Filter & Magnitude range & $d\log N/dm$ \\ 
\\
B& 20.0-24.0 & $0.45\pm0.01$\\
B& 24.0-25.5 & $0.32\pm0.04$\\
V& 20.0-24.0 & $0.40\pm0.03$\\
R& 20.0-24.0 & $0.37\pm0.03$\\
I& 20.0-24.0 & $0.34\pm0.02$\\

\end{tabular}     
\caption{Best-fitting galaxy count slopes in $BVRI$ bands for our
  catalogue, computed in the specified magnitude ranges.}
\label{tab:f02:slopes}
\end{table}
\nocite{2001AJ....122.1104Y}
\nocite{2001A&A...379..740A}

\subsection{Stellar colours}
\label{sec:comp-stell-colo}

In this Section we examine the relative and absolute photometric
calibration by investigating the colours of stars distributed over our
field.

\begin{figure}
\resizebox{\hsize}{!}{\includegraphics{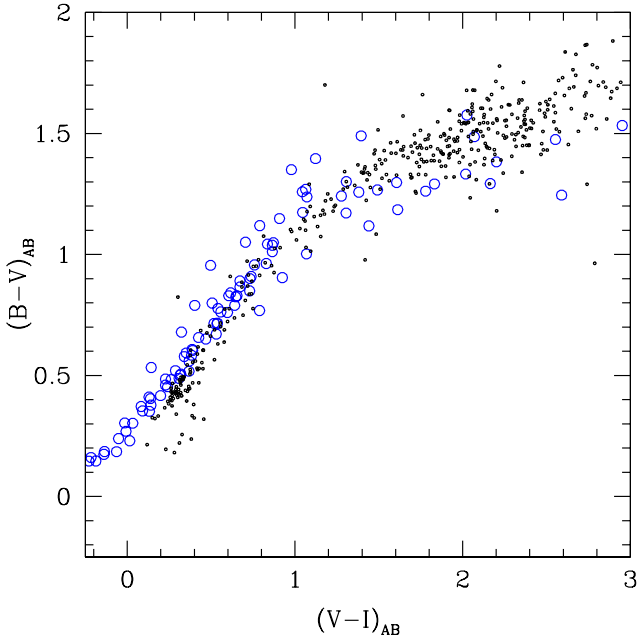}}
\caption{$(B-V)_{AB}$ vs $(V-I)_{AB}$ colours for  (filled points)
  compared to synthetic colours computed from library of Pickles (open
  circles). }
\label{fig:bvvi_fig1}
\end{figure}

The filled symbols in Figure~\ref{fig:bvvi_fig1} shows the $(B-V)_{AB}$
and $(V-I)_{AB}$ colours of point-like sources in our catalogue
compared to synthetic colours generated by convolving the
stellar spectral energy distributions of \cite{1998PASP..110..863P} with
the CFH12K filter set (open symbols).  From this plot it is apparent
that in $BVI$ filters our absolute calibration is accurate to within
$\sim 0.05$ magnitudes.

\begin{figure}
\resizebox{\hsize}{!}{\includegraphics{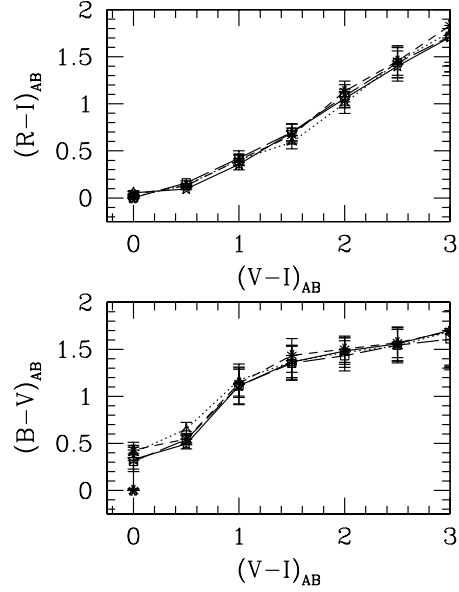}}
\caption{Stellar colours as a function of position for each of the four
  pointings. In each pointing we compute the median $(B-V)_{AB}$ and
  $(R-I)_{AB}$ colours in bins of $(V-I)_{AB}$ (upper and lower panels
  respectively). This is represented as the four different line styles.
  The regions are defined as follows: solid line: south-west, dotted
  line: north-east; dashed line: south-east, dot-dashed line:
  north-west.}
\label{fig:vi_med_all}
\end{figure}

We also use measured colours of stars in our fields to investigate if
there are systematic differences in stellar colours across the mosaic.
The results of these tests are illustrated in
Figures~\ref{fig:vi_med_all} where we plot the median $(B-V)_{AB}$,
$(V-I)_{AB}$ and $(R-I)_{AB}$ colours of stars selected from each of
the four pointings. It appears from these tests that over these four
regions the relative differences in zero points is less than $0.05$
magnitudes for each of the four filters.

\subsection{Field galaxy colours}
\label{sec:field-galaxy-colours}

\begin{figure}
\resizebox{\hsize}{!}{\includegraphics{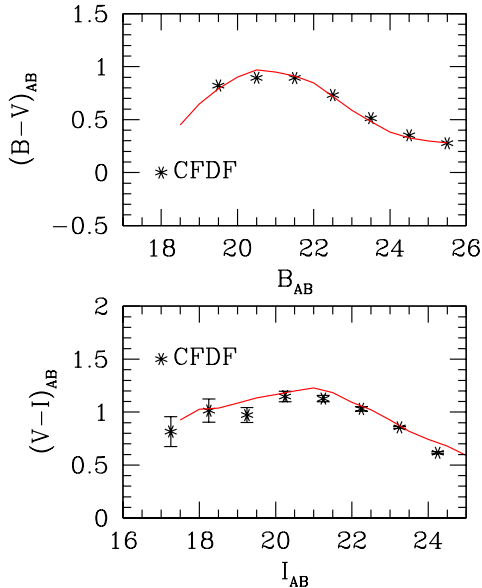}}
  \caption{Field galaxy colours from our catalogues. In the
    upper panel, the solid line shows the median $(B-V)_{AB}$ galaxy
    colour computed in half-magnitude bins as a function of $B_{AB}$
    limiting magnitude; in the lower panel, we show the median
    $(V-I)_{AB}$ colour as a function of $I_{AB}$~magnitude. In both
    cases we compare with colours from the CFDF survey (symbols). }
\label{fig:BVVIcol}
\end{figure}

In this Section we compare our field galaxy colours (or, more
accurately, extended sources) to those from other deep surveys. We
compute the median $(B-V)_{AB}$ and $(V-I)_{AB}$ colours in
half-magnitude bins as a function of $B_{AB}$ and $I_{AB}$ sample
limiting magnitude. These measurements are shown as the solid lines in
the upper and lower panels of Figure~\ref{fig:BVVIcol}. The symbols
show field galaxy colours measured from the CFDF survey. In both cases
our agreement with the published values is excellent.

\begin{figure}
\resizebox{\hsize}{!}{\includegraphics{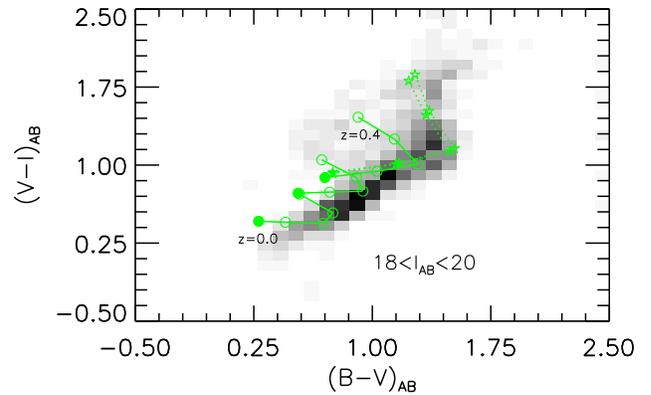}}
  \caption{$(B-V)_{AB}$ vs $(V-I)_{AB}$ colour-colour diagram for 
    galaxies with $18<I_{AB}<20$. The greyscale levels are proportional
    to the numbers of galaxies at each combination of $(V-I)_{AB}$ and
    $(B-V)_{AB}$ colour. The lines shows model tracks for five galaxy
    types (two early types, shown by the dotted lines, and three
    late-types, represented by the solid lines). Tracks are computed in
    the redshift range $0.0<z<0.4$ in intervals of $0.1$. Solid
    symbols show the location of the $z=0$ point for each track. }
\label{fig:BVVIcol_1820}
\end{figure}

\begin{figure}
\resizebox{\hsize}{!}{\includegraphics{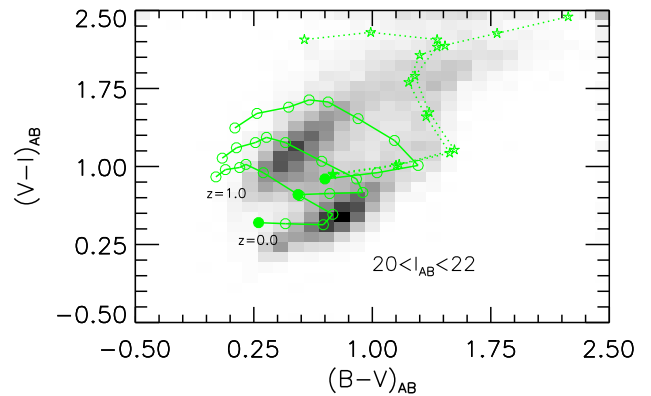}}
  \caption{Colour-colour diagram for galaxies with $20<I_{AB}<22$. For
    this Figure, model tracks span the redshifts range $0.0<z<1.0$. }
\label{fig:BVVIcol_2022}
\end{figure}

\begin{figure}
\resizebox{\hsize}{!}{\includegraphics{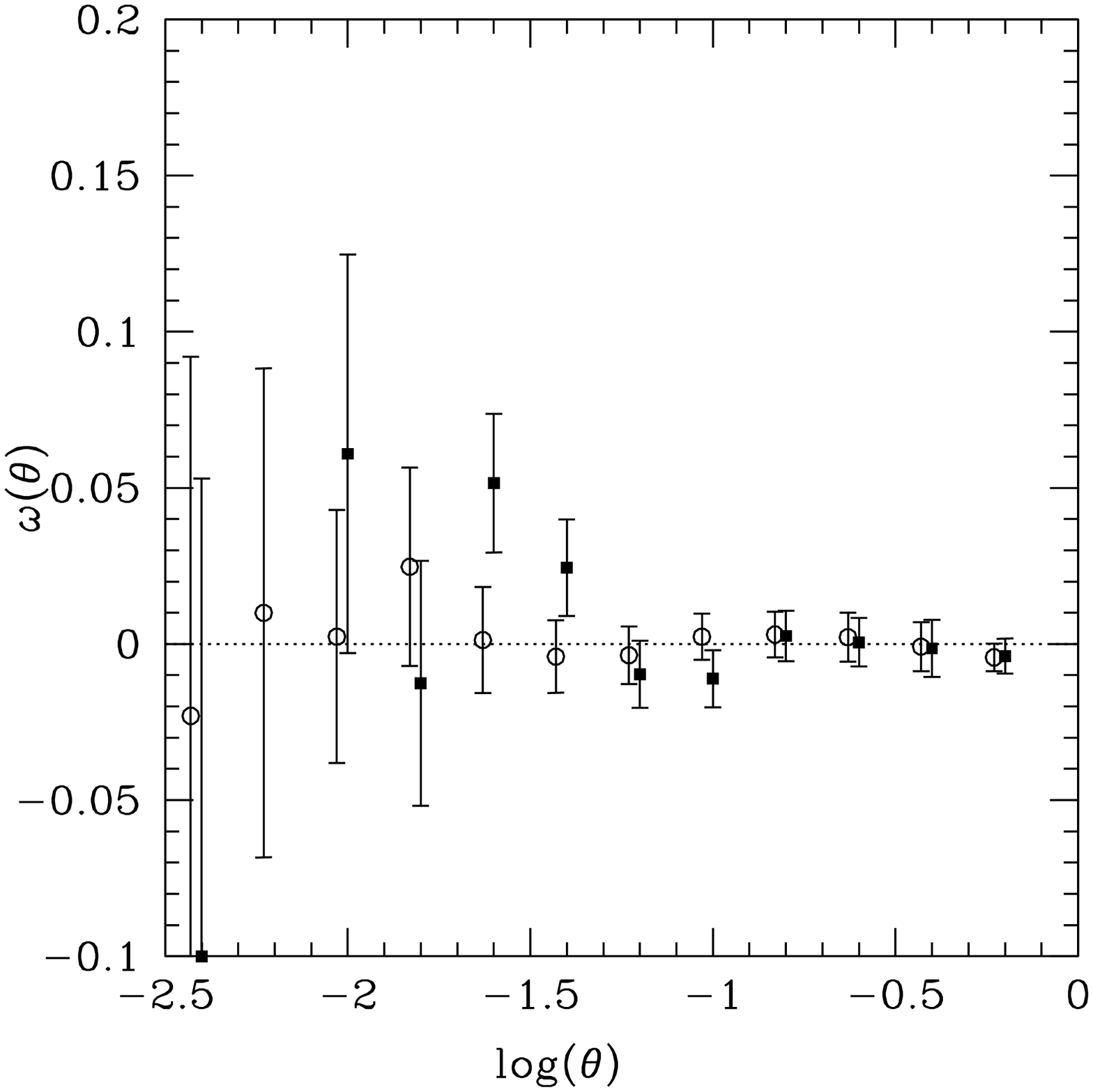}}
  \caption{Colour-colour diagram for galaxies with
    $22<I_{AB}<24$. Model tracks are computed from $0.0<z<3.0$.}
\label{fig:BVVIcol_2224}
\end{figure}

We also investigate the distribution of galaxies in colour-colour space
as a function of magnitude; this is illustrated in
Figures~\ref{fig:BVVIcol_1820},~\ref{fig:BVVIcol_2022} and
\ref{fig:BVVIcol_2224}. In these Figures we compare our observed galaxy
colours to model predictions for a range of progressively fainter
magnitude slices. The grey-scales in these Figures are proportional to
object number density, computed in 0.1 magnitude intervals in
$(B-V)_{AB}$ and $(V-I)_{AB}$. Dotted lines show the path of late-type
(E, S0) galaxies wheras early-type galaxies (Sa, Sc and Irregular) are
represented by solid lines.  Galaxy colours were computed at intervals
of $0.1$ in redshift by applying the CFH12K filter response functions
to model evolving spectral energy distributions generated using the
``2000'' revision of the GISSEL libraries \citep{BC}. Computations were
made assuming a flat $\Lambda$-CDM cosmology. The e-folding times for
the five tracks are $(0.3,1.0,2.0,4.0)$ Gyrs and ``constant'' for
elliptical, S0, Sa, Sc and Irregular galaxy types respectively.  Model
galaxy zero-redshift colours (represented by the filled dots) are in
good agreement with the colours reported in \cite{2001MNRAS.326..745M}.
Finally, intergalactic absorption as modelled by
\cite{1995ApJ...441...18M} has also been added.

Figure~\ref{fig:BVVIcol_1820} shows the bright magnitude slice at
$18<I_{AB}<20$. In this Figure, most galaxies lie in a single
well-defined region. From previous $I_{AB}$-selected galaxy redshift
surveys such as the CFRS we know that at these magnitudes the median
redshift of the faint galaxy population is around $z\sim0.2$
\citep{1995ApJ...455...96C}, and this is also roughly where the model
tracks enter the locus occupied by the galaxy population. In addition,
we note that the model tracks span very well all the observed colours
in our survey.

For the intermediate magnitude slice at $20<I_{AB}<22$, illustrated in
Figure~\ref{fig:BVVIcol_2022}, the (predominantly late-type) galaxy
population occupies two distinct locii, and in colour-colour space
there is a comparatively well-defined separation between high ($z>0.4$)
and low ($z<0.4$) redshift galaxies.  For the very faintest magnitude
slice, $22<I_{AB}<24$, the sharp distinction has been removed, and all
parts of the colour-colour diagram are equally populated. In general,
our extended source colours are in agreement with the existing model
tracks. 

\subsection{Clustering of point-like sources}
\label{sec:clust-stell-sourc}

In this Section and the following one we investigate the clustering
properties of point-like and extended sources in our catalogues. To do
this, we use the well-known projected two-point angular correlation
function, $\omega(\theta)$, which measures the excess of pairs
separated by an angle $\theta, \theta+\delta\theta$ with respect to a
random distribution. This statistic is useful for our purposes because
it is particularly sensitive to any residual variations of the
magnitude zero-point across our stacked images.

We first measure the angular correlation function $\omega(\theta)$ of
the stellar sources.  As stars are unclustered, we expect that if our
magnitude zero-points and detection thresholds are uniform over our
field then $\omega(\theta)$ should be zero at all angular scales.  For
this test, we consider two magnitude limited samples: point-like
sources with $18.0<I_{AB}<20.0$ and those with $20.0<I_{AB}<21.5$. We measure
$\omega(\theta)$ using the standard \cite{LS} estimator, i.e.,

\begin{equation}
\omega ( \theta) ={\mbox{DD} - 2\mbox{DR} + \mbox{RR}\over \mbox{RR}}
\label{eq:1.ls}
\end{equation}

with the $DD$, $DR$ and $RR$ terms referring to the number of
data--data, data--random and random--random galaxy pairs between
$\theta$ and $\theta+\delta\theta$. We use logarithmically spaced bins
with $\Delta\log(\theta)=0.2$. 

\begin{figure}
\resizebox{\hsize}{!}{\includegraphics{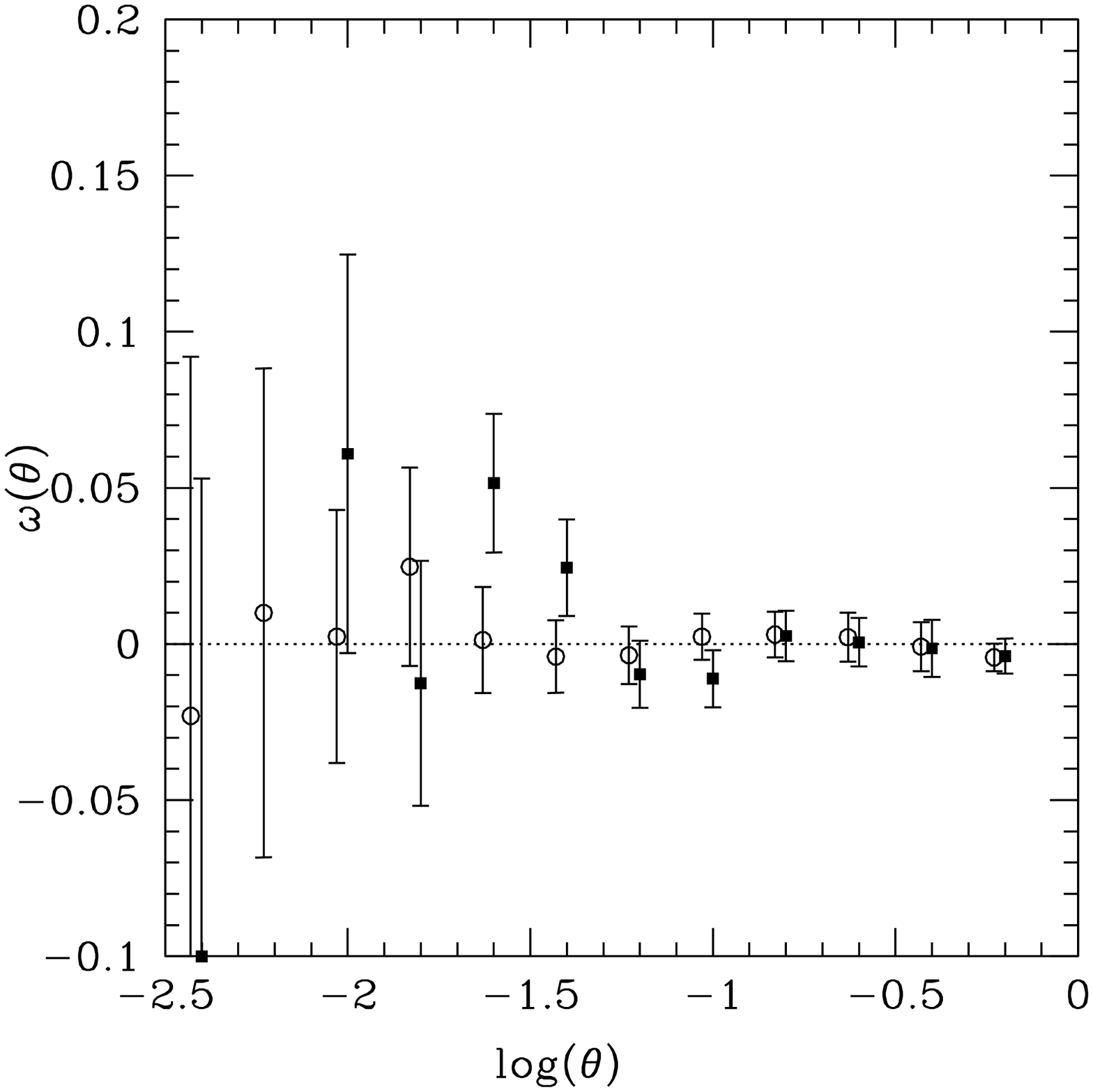}}
\caption{Angular clustering of stellar sources, $\omega(\theta)$ as a
  function of the logarithm of the angular separation in degrees.
  Filled squares represent objects with $18.0<I_{AB}<20$; open circles
  show sources with $20.0<I_{AB}<21.5$. For clarity, the fainter
  magnitude slice has been slightly offset in $\log(\theta)$ from the
  brighter one.}
\label{fig:wigstars}
\end{figure}

Our results are displayed in Figure~\ref{fig:wigstars}. For both
magnitude cuts, at scales $-2.5<\log(\theta)<0.0$ the measured correlation
amplitudes are consistent with zero.

\subsection{Clustering of extended sources}
\label{sec:clust-gal-sourc}

The angular clustering of field galaxies, parametrised by
$\omega(\theta)$ provides important information concerning the galaxy
distribution.  Many studies have measured $\omega(\theta)$ for
progressively larger samples of galaxies reaching to fainter and
fainter magnitudes (see, for example, \citet{BSM,RSMF,EBK}). The advent
of mosaic detector arrays means it is now possible to measure the
angular clustering of field galaxies on degree scales at $z\sim1$ with
samples reaching magnitude limits of $I_{AB}\sim24$. Both
\citet{2001A&A...376..756M} and \citet{2003ApJ...585..191W} have both
measured $\omega(\theta)$ using deep catalogues constructed using the
UH8K camera. In this Section we present comparisons between
$\omega(\theta)$ measurements in this survey with other published works
as a further check on the reliability of our data reduction procedure.

\begin{figure}
\resizebox{\hsize}{!}{\includegraphics{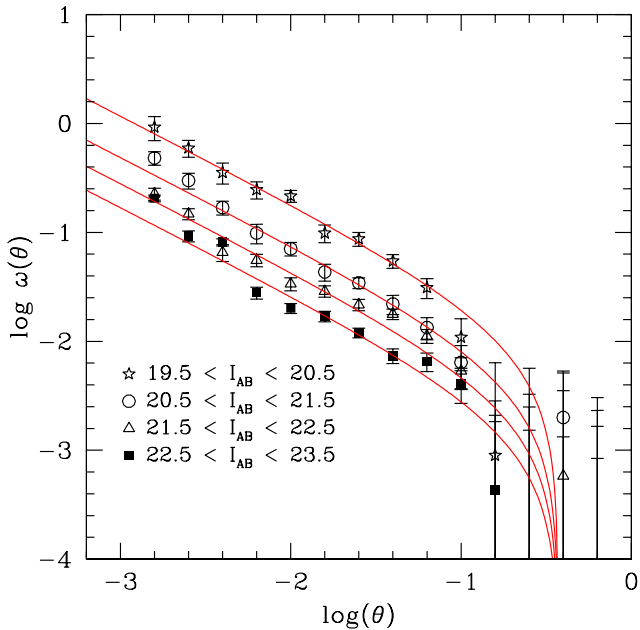}}
\caption{$I_{AB}$-selected galaxy clustering. 
  For each magnitude slice, we show the amplitude of the angular
  correlation function, $\omega(\theta)$ as a function of the logarithm
  of the angular separation in degrees, $\log(\theta)$.  At each
  magnitude slice, we show the fitted correlation amplitude, including
  a correction for the integral constraint term. }
\label{fig:wiggals}
\end{figure}

In Figure~\ref{fig:wiggals} we show the angular correlation function
$\omega(\theta)$ as a function of angular separation, $\log(\theta)$
for $19.5<I_{AB-med}<23.5$ measured from our catalogues.  We compute
$\omega(\theta)$ for $-2.8<\log(\theta)<0.0$ . The solid lines show the
fitted correlation amplitudes assuming a functional form
$\omega(\theta)=A\theta^{-0.8}-C$ parametrisation.  The ``C'' term
represents the integral constraint correction (see, for example,
\cite{RSMF}). For our field, we find that $C=3.0$. Measurements in our
catalogue follow the expected power-law behaviour at least until
angular scales where the integral constraint correction becomes
important.

\begin{figure}
\resizebox{\hsize}{!}{\includegraphics{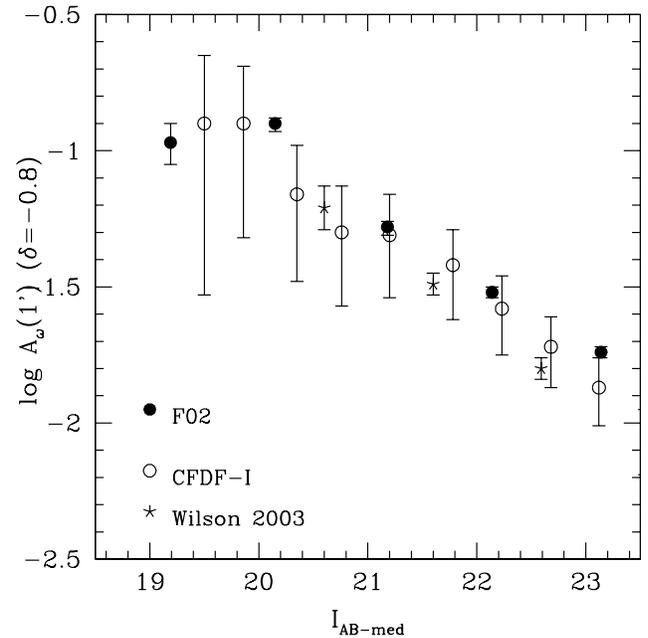}}
\caption{The amplitude of the angular correlation function at 
  arcminute scales. The filled symbols show $\log A_\omega(1\arcmin)$
  as a function of $I_{AB}$ limiting magnitude as measured in the
  CFH12K-VIRMOS deep field, assuming a power-law form for the galaxy
  angular correlation function $\omega(\theta)$ with slope of
  $\delta=-0.8$ (error bars shown represent only the error bars from
  the fitting program and do not indicate the true cosmic errors, which
  are likely to be larger). The open circles and stars show
  measurements from other recent wide-field surveys.}
\label{fig:wigmag}
\end{figure}

In Figure~\ref{fig:wigmag} we show the amplitude of $\omega(\theta)$ at
$1\arcmin$ (assuming a slope $\delta=-0.8$) a function $I_{AB}$ sample
median magnitude for the same magnitude slices presented in
Figure~\ref{fig:wiggals}. The open symbols and stars show the
measurements from the CFDF survey and \citet{2003ApJ...585..191W}
respectively. The amplitude of $\omega(\theta)$ at $1\arcmin$ steadily
decreases with limiting magnitude: moreover, all magnitude bins agree
quite well with the literature measurements.

It has recently reported that CFH12K observations may be affected by
scattered
light\footnote{http://www.cfht.hawaii.edu/Instruments/Elixir/scattered.html}.
This effect originates from light reflecting from structures inside the
camera/telescope assembly. It produces an additive correction which is
not removed by flat-fielding, as the flat-fields themselves are not
uniformly illuminated.  Measurements made at CFHT (and described
extensively on the ELIXIR web pages) show that object magnitudes
measured in the outer edges of the field of view can be as much 0.1
magnitudes fainter than in the central region.  

We have looked for possible signatures of this problem in our
catalogues.  At large angular scales, the stellar correlation function
we measure (Section~\ref{sec:clust-stell-sourc}) is consistent with
zero (Figure~\ref{fig:wigstars}). However, because the slope of the
stellar number counts is relatively shallow compared to the galaxy
counts (Section~\ref{sec:galaxycounts}) the radial dependence of
stellar counts will be comparatively weaker (corresponding to density
changes of around a few percent). For galaxies, the steeper count slope
means that the radial density dependence is higher, and so would be
easier to detect this with a correlation function analysis as outlined
above.

We used a scattered light image supplied by CFHT to investigate if the
magnitude variation can have a significant effect on the measured
$\omega(\theta)$. Because of the small dither offsets used in our
observation strategy we are able to separate a single pointing from the
four-pointing combined mosaic, and from this extracted sub-area we
'correct' magnitudes in our catalogues based on the radial dependence
derived at CFHT, and then re-apply our selected magnitude cuts. For
galaxies with $18.5<I_{AB}<24.5$ we find that application of the
scattered light map produces a difference of around $\sim 3.5\%$ in the
total numbers of galaxies (below the systematic errors quoted in the
previous sections). Furthermore, the measured $\omega(\theta)$ values
computed at each angular separation do not differ significantly before
and after the application of the scattered light correction (the
difference is smaller than the Poissonian error bar in each bin). From
these considerations, we conclude that the effects of residual
scattered light in our catalogues is not detectable, at least to
$I_{AB}<24.0$.

\section{Summary}
\label{sec:summary}

In this paper we have presented $BVRI$ optical data for the
CFH12K-VIRMOS F02 deep field. Our observations reach limiting
magnitudes of $B_{AB}\sim 26.5$, $V_{AB}\sim26.2$, $R_{AB}\sim25.9$ and
$I_{AB}\sim25.0$ (measured from our simulations). This field
consists of a single contiguous area covering $1.2$~deg$^2$. In the
magnitude range $18.0<I_{AB}<24.0$ there are 90,729 extended sources in
our survey.

We have presented a detailed explanation of the reduction and
preparation of the object catalogues and described how our astrometric
and photometric solutions were computed. In summary, $68\%$ of
catalogue sources have an absolute per co-ordinate residual less than
$\vert\Delta\alpha\vert\sim0.38\arcsec$ and
$\vert\Delta\delta\vert\sim0.32\arcsec$ in right ascension and
declination respectively. Expressed in the same sense, our internal
(filter-to-filter) per co-ordinate astrometric uncertainties are
$\vert\Delta\alpha\vert\sim0.08\arcsec$ and
$\vert\Delta\delta\vert\sim0.08\arcsec$. Our absolute photometric
calibration, estimated from standard star data, and from galaxy number
counts, is accurate to $\sim0.1$~magnitudes. A series of
Monte-Carlo simulations have been used to assess the reliability of
catalogues in planes defined by peak surface brightness, position, and
total magnitude.  These show that our catalogues are reliable, uniform
and fully complete to $I_{AB}<24.0$ .

Galaxy number counts in $BVRI$ bandpasses have been presented. We
describe a method of star-galaxy separation which correctly accounts
for the position-dependent seeing present in our final stacked images.
Our galaxy counts are in good agreement with literature compilations,
and our stellar counts are well matched by a theoretical model of the
galaxy. Qualitatively, at each magnitude bin, $I_{AB}$ selected galaxy
number counts in our survey agree to within $10\%$ with measurements
made in the CFDF survey. Additionally, the slope of the galaxy counts
in both $B$ and $I$ are consistent with previous determinations from
other deep surveys.  Furthermore, field galaxy colours in $(V-I)$ and
$(B-V)$ measured in our survey are also in agreement (better than
$0.1~$mag) with measurements made in the Canada-France deep fields
survey. The location of galaxies in colour-colour space, and its
evolution with magnitude, is broadly consistent with expectations based
on simple populations synthesis models.

We have also measured the angular correlation function of both stars
and galaxies in our dataset. Stellar correlations are consistent with
zero at all angular scales. Clustering of galaxies measured to
$I_{AB-med}\sim23$ in the angular range $-3<\log(\theta)<-1$ follows
well the expected power law behaviour, and the observed scaling of the
amplitude of $\omega(\theta)$ with $I_{AB-med}$ is in agreement with
previous surveys. We have also investigated the possible effect that
scattered light could have on the galaxy angular correlation function,
and concluded that to $I_{AB}<24.0$ this effect does not pose a
significant problem to our measurements. 

Future papers will describe numerous extensions in wavelength to the
survey field described here. Furthermore the catalogues presented in
this paper are currently being used to select a very large
magnitude-limited sample of objects for the VIRMOS-VLT deep survey.
Together, both surveys will provide a unique picture of the Universe at
intermediate and high redshifts.
 
\begin{acknowledgements}
  HJMCC acknowledges the use of TERAPIX computer facilities at the
  Institut d'Astrophysique de Paris, where much of this work was
  carried out. We thank Lucia Pozzetti for help with the evolutionary
  tracks in Section~\ref{sec:field-galaxy-colours}. We also to
  acknowledge assistance from the members VIRMOS consortium, and an
  anonymous referee.  HJMCC's work has been supported by MIUR
  postdoctoral grants COFIN-00-02, COFIN-00-03 and a VIRMOS
  postdoctoral fellowship. Y.M., E.B. and M.R. were partly funded by
  the European RTD contract HPRI-CT-2001-50029 "AstroWise".
\end{acknowledgements}

\bibliographystyle{apj}
%\bibliography{aamnemonic.bib,thesis}

\end{document}